\documentclass[
superscriptaddress,
twocolumn,
amsmath,
amssymb,
aps,
prl
]{revtex4-2}
\usepackage{bm}
\usepackage{graphicx}
\usepackage[colorlinks=true,
       	    linkcolor=blue,
       	    citecolor=blue,
       	    filecolor=blue,
       	    urlcolor=blue,
            ]{hyperref}

\begin{document}

\title{Linearly-polarized Coherent Emission from Relativistic Magnetized Ion-electron Shocks}

\author{Masanori Iwamoto}
\email{masanori.iwamoto@yukawa.kyoto-u.ac.jp}
\affiliation{Yukawa Institute for Theoretical Physics, Kyoto University, Kitashirakawa-Oiwakecho, 
Sakyo-Ku, Kyoto 606-8502, Japan}
\affiliation{Faculty of Engineering Sciences, Kyushu University, 6-1, Kasuga-koen, Kasuga, Fukuoka, 816-8580, 
Japan}

\author{Yosuke Matsumoto}
\affiliation{Institute for Advanced Academic Research, Chiba University, 1-33 Yayoi, Inage-ku, Chiba, Chiba 263-8522, Japan}

\author{Takanobu Amano}
\affiliation{Department of Earth and Planetary Science, University of Tokyo,
7-3-1 Hongo, Bunkyo-ku, Tokyo 113-0033, Japan}

\author{Shuichi Matsukiyo}
\affiliation{Faculty of Engineering Sciences, Kyushu University, 6-1, Kasuga-koen, Kasuga, Fukuoka, 816-8580, 
Japan}

\author{Masahiro Hoshino}
\affiliation{Department of Earth and Planetary Science, University of Tokyo,
7-3-1 Hongo, Bunkyo-ku, Tokyo 113-0033, Japan}

\begin{abstract}

Fast radio bursts (FRBs) are millisecond transient astrophysical phenomena and 
bright at radio frequencies. The emission mechanism, however, remains unsolved 
yet. One scenario is a coherent emission associated with the magnetar flares and 
resulting relativistic shock waves. Here, we report unprecedentedly large-scale
simulations of relativistic magnetized ion-electron shocks, showing that strongly 
linear-polarized electromagnetic waves are excited. 
The kinetic energy conversion to the emission is so efficient that the wave amplitude 
is responsible for the brightness. We also find a polarization angle swing 
reflecting shock front modulation, implicating the polarization property of some 
repeating FRBs. The results support the shock scenario as an origin of the FRBs.

\end{abstract}

\maketitle

Fast radio bursts (FRBs) are luminous millisecond-duration pulses detected at 
radio frequencies near 1 gigahertz (GHz) mostly from extragalactic origins \citep{Lorimer2007,Petroff2022}.
Some FRBs are known to repeat, while most of them do not.
The mechanism powering the non-repeating FRBs remains a topic of debate 
\citep{Katz2022b,Bhandari2023}. On the other hand,
magnetars are often invoked for the progenitor of the repeating FRBs \citep{Lyubarsky2021},
which is supported by the recent discovery of FRB 200428 
associated with a galactic magnetar \citep{Andersen2020,Bochenek2020}.
The extremely high brightness temperature of FRBs requires coherent emission in 
the sense that electron bunches collectively emit electromagnetic waves \citep{Katz2014}. 
One of the promising coherent emission mechanisms
is synchrotron maser instability (SMI) in relativistic magnetized shocks induced by the magnetar flares 
\citep{Lyubarsky2014,Beloborodov2017,Beloborodov2020,Metzger2019,Margalit2019,Margalit2020}.
The fundamental properties of the SMI have been studied by using ab-initio Particle-in-Cell (PIC) simulations 
and have been confirmed that the coherent emission is intrinsic to relativistic magnetized shocks 
\citep{Langdon1988,Hoshino1991, Hoshino1992,Gallant1992,Amato2006,Sironi2011,Iwamoto2017,Iwamoto2018,Plotnikov2019,Sironi2021}. 
The SMI in the context of relativistic magnetized shocks can self-consistently convert the incoming flare energies 
into coherent emission.

The observed rotation measure of some repeating FRBs indicates the magneto-ionic environments of the sources \citep{Mckinven2023},
and thus relativistic magnetized shocks can be induced in baryon-loaded shells \citep{Metzger2019,Margalit2019,Margalit2020}.
Although the SMI model usually assumes the energy conversion ratio from incoming total energy into 
electromagnetic wave energy $f_{\xi} \sim 10^{-3}$, which was confirmed by PIC simulations of pair (electron-positron) shocks 
\citep{Iwamoto2017,Iwamoto2018,Plotnikov2019}, $f_{\xi}$ in ion-electron shocks remains unclear especially in 
realistic three-dimensional (3D) systems.
The observational fact that repeating FRBs often exhibit the high degree of linear polarization \citep{Masui2015,Michilli2018,Petroff2019,Luo2020,Day2020}
constrains the emission mechanism as well. Previous 2D PIC simulations demonstrate the excitation of the two 
linearly polarized waves: extraordinary (X) and ordinary (O) mode waves \citep{Iwamoto2018,Ligorini2021a,Ligorini2021b}.
3D shock simulations are required for properly taking into account the both X and O mode wave contribution
to the polarization.
In this Letter, we demonstrate that $f_{\xi} \sim 10^{-3}$ is indeed satisfied and 
determine the precise state of the polarization based on the Stokes parameter analysis.
Our unprecedentedly large-scale PIC simulations of 3D ion-electron shocks reveal the underlying physical 
process of the SMI-induced coherent emission and provide the detailed description 
of the wave properties.

We quantify the emission efficiency and polarization properties of the coherent emission by using a fully kinetic electromagnetic 
PIC code, which enables long-term stable calculations of a relativistic plasma flow 
\citep{Matsumoto2015,Matsumoto2017,Ikeya2015}. The Japanese flagship supercomputer Fugaku at the RIKEN Center for Computational 
Science helps us to perform 3D ion-electron shock simulations.
The unit of length is the electron skin depth $c/\omega_{pe}$,
which is the characteristic electron kinetic scale and resolved with 20 computational 
cells. The simulation time step is set as $0.05\omega_{pe}^{-1}$. 
In the above expression, $c$ is the speed of light and 
$\omega_{pe}=\sqrt{4\pi N_1 e^2/\gamma_1 m_e}$ is 
the relativistic electron plasma frequency with the upstream electron number density 
$N_1$ and bulk Lorentz factor of the upstream plasma flow $\gamma_1$.
The computational domain is a square prism with $0 \leq x/(c/\omega_{pe}) \leq 2000$, 
$0 \leq y/(c/\omega_{pe}) \leq 46$, and  $0 \leq z/(c/\omega_{pe}) \leq 46$ for $\sigma_i=0.1$
and  $0 \leq x/(c/\omega_{pe}) \leq 2000$, 
$0 \leq y/(c/\omega_{pe}) \leq 23$, and  $0 \leq z/(c/\omega_{pe}) \leq 23$ for $\sigma_i=0.5$,
where $ \sigma_i=B_1^2/4\pi \gamma_1 N_1 m_i c^2$ is the ion magnetization parameter and
$\bm{B_1}=(0,0,B_1)$ is the upstream background magnetic field.  
The periodic boundary condition is applied in the $y$ and $z$ directions 
for both particles and fields.
The lower $x$ boundary at $x=0$ is the conducting wall.
A cold ion-electron plasma flow drifting in the $-x$ direction 
are continuously injected from the upper $x$ boundary.
The interaction between the injected and reflected plasma flow triggers shocks propagating 
the $+x$ direction, and thus the simulation frame corresponds to the downstream rest frame.
We examine the two cases $\sigma_i=0.1$ and $0.5$, which are motivated by the SMI model 
\citep{Metzger2019,Margalit2019,Margalit2020}.
The ion-to-electron mass ratio is fixed as $m_i/m_e=200$ throughout this study.
Note that the electron magnetization parameter $\sigma_e=m_i \sigma_i/m_e =20$ and $100$ is satisfied 
and electrons are highly magnetized. 
We consider the highly relativistic plasma flow with $\gamma_1=40$. 
The number of particles per electron skin depth per species in the upstream is set as 
$N_1(c/\omega_{pe})^3=32000$. 
The transverse box size is comparable to the upstream ion gyro-radius, 
which is sufficiently large to capture essential physical processes reported by previous 
2D simulations under different parameters and configurations
\citep{Iwamoto2017,Iwamoto2018,Iwamoto2019}.

\begin{figure*}[htb]
	\includegraphics[width=16cm]{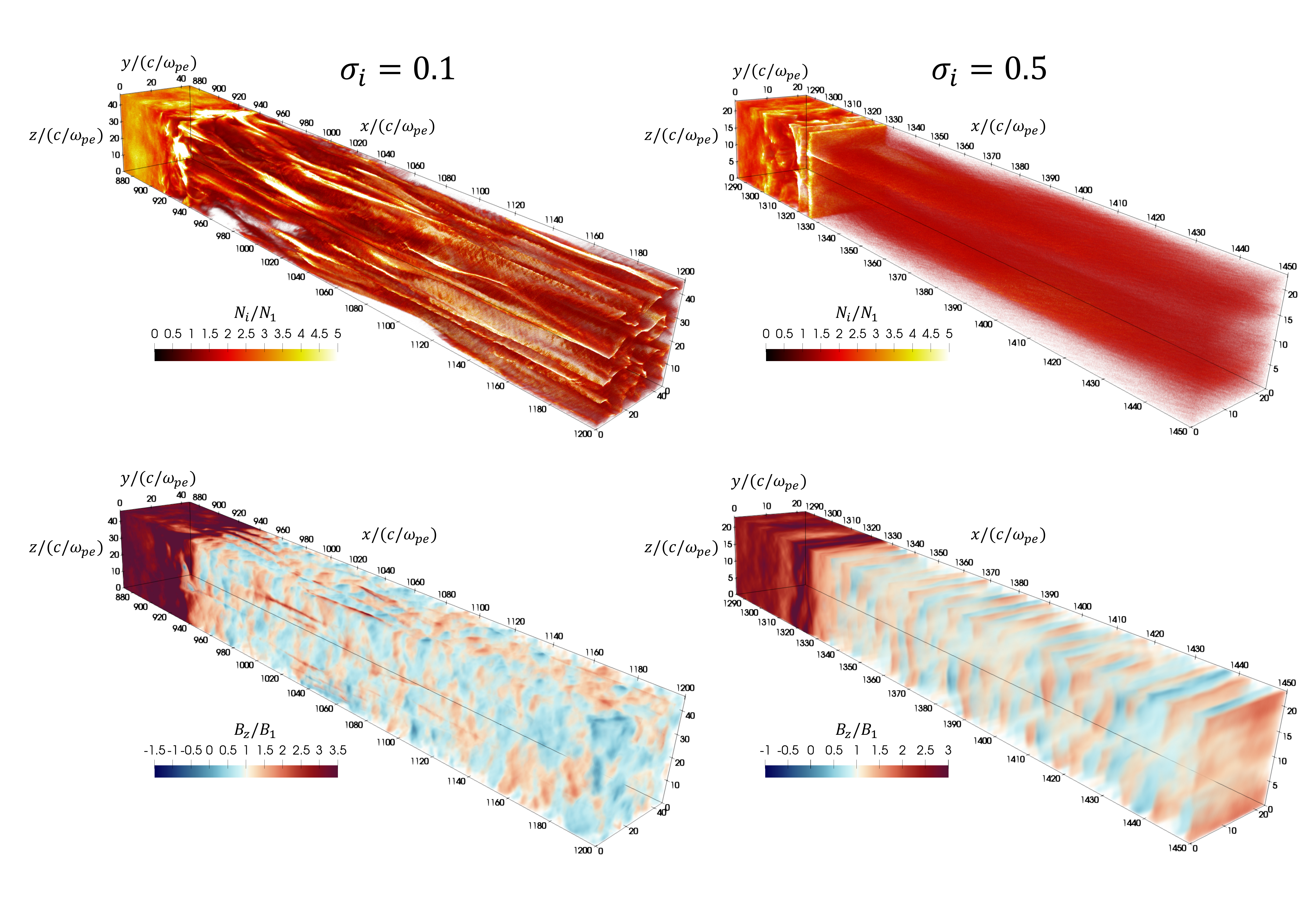}
	\caption{Global structures of relativistic magnetized shocks. The ion number density (top) and $z$ components of magnetic 
	field (bottom) at the final state $\omega_{pe} t=2000$ are shown for $\sigma_i=0.1$ (left) and $0.5$ (right). The color scales 
	for the two cases of $\sigma_i$ are the same.}
	\label{fig:3dshock}
   \end{figure*}

Figure \ref{fig:3dshock} shows the snapshots for $\sigma_i=0.1$ (left) and $0.5$ (right) at the final state of our simulations $\omega_{pe} t=2000$. 
All physical quantities are normalized by the corresponding upstream ones. The ion number density $N_i$ (top) is strongly modified 
in the upstream region due to the filamentation instability (FI). The FI is a transverse self-modulation of an electromagnetic wave 
\citep{Sobacchi2020,Sobacchi2022,Ghosh2022,Sobacchi2023,Iwamoto2023} and the density filaments are also observed in the previous 
simulations of 2D ion-electron shock \citep{Iwamoto2019,Ligorini2021a,Ligorini2021b}. $N_i$ for relatively high magnetization
$\sigma_i=0.5$ exhibits sheet-like structures perpendicular to the ambient magnetic field rather than filamentary structures, 
which is consistent with the previous simulations of 3D pair shocks with high magnetization \citep{Sironi2021}. 
This is probably because the magnetic pressure dominates over the ponderomotive force for high 
magnetization and particles are preferentially pushed along the ambient magnetic field \cite{Sobacchi2020,Sobacchi2022}. 
The magnetic field $B_z$ (bottom) show large-amplitude electromagnetic waves are excited by the SMI. 
Since the ponderomotive force exerted by the electromagnetic waves induces the FI, 
the FI gets weaker as the wave amplitude gets smaller \citep{Sobacchi2023,Iwamoto2023}.
As will be shown later,
the radiant power for $\sigma_i=0.5$ is smaller than that for $\sigma_i=0.1$.
Therefore, the FI for $\sigma_i=0.5$ is relatively weak and the wave propagation is not strongly disturbed.
We thus think that the electromagnetic waves for $\sigma_i=0.5$ are mostly planar.
The filamentary structures 
corresponding to the density filaments are seen for $\sigma_i=0.1$. The electromagnetic waves are accumulated in 
the low density region $N_i/N_1 < 1$, indicating that the dispersion measure of FRBs can be modified \citep{Sobacchi2023}. 
The fluctuation of the inferred dispersion measure \citep{Katz2022a} can be attributed to the FI via the SMI.

\begin{figure}[htb]
	\includegraphics[width=8.5cm]{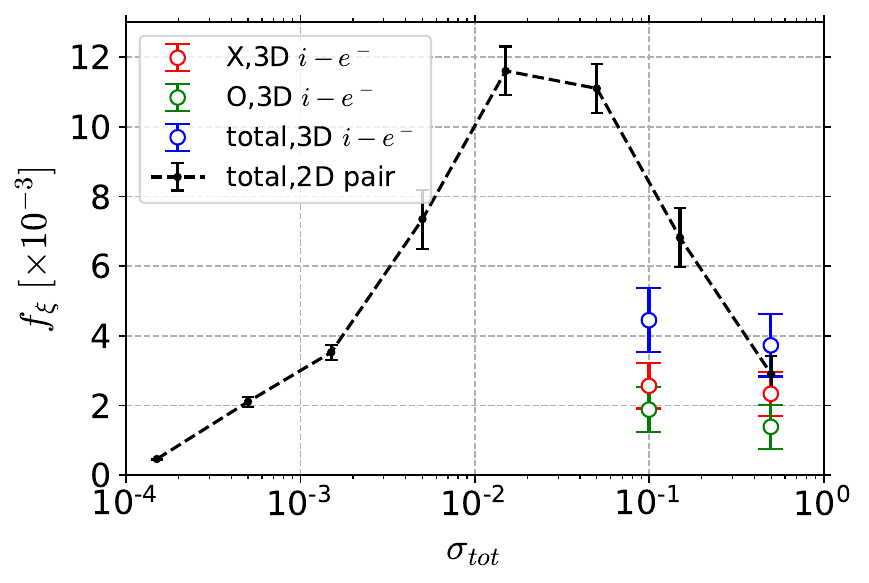}
	\caption{Energy conversion ratios as a function of total magnetization determined from the 
	snapshots at the final state $\omega_{pe}t=2000$. The X, O, total waves are shown in red, green, 
	and blue circles, respectively. The black dashed line indicates the total power measured in previous 
	2D pair shock simulations for comparison \citep{Iwamoto2018}.}
	\label{fig:fxi}
\end{figure}

The energy conversion ratio $f_{\xi}$ is quantified as a function of total magnetization 
$\sigma_{tot}=\sigma_i/(1+m_e/m_i)\simeq \sigma_i$ in the downstream rest frame and shown in Figure \ref{fig:fxi} with circles.
The conversion ratios from the incoming total energy to the X mode, O mode, total (i.e., X+O) wave energies 
are shown in red, green, and, blue, respectively.
We determine $f_\xi$ in the upstream region 
$1400 \leq x/(c/\omega_{pe}) \leq 1700$ for $\sigma_i=0.5$ and $1000 \leq x/(c/\omega_{pe}) \leq 1300$ for $\sigma_i=0.1$ at 
$\omega_{pe} t=2000$. The X mode wave amplitude is systematically larger than the O mode wave because the O mode waves are induced 
by the magnetic field fluctuations along the ambient magnetic field due to the Alfven ion cyclotron instability (AIC) which are 
suppressed for high magnetization \citep{Iwamoto2018}. Note that the linear theory of 
the SMI predicts only the X mode wave excitation \citep{Hoshino1991}. 
Present results show that the assumption $f_{\xi} \sim 10^{-3}$ is valid even 
for 3D ion-electron shocks. The black dashed line indicates the previous simulation results of 
2D pair shocks \citep{Iwamoto2018}.
Note that the emission efficiency in pair shocks can be twice as high as ion-electron shocks because both electrons and positrons 
contribute to the electromagnetic wave emission via the SMI. 
The emission efficiency of 2D pair shocks is almost comparable to that of 3D ion-electron shocks due to the ion-electron coupling 
\citep{Lyubarsky2006,Hoshino2008,Iwamoto2019}. 
Since $f_{\xi}$ in pair shocks remains unchanged in 2D and 3D for $\sigma_{tot} \gtrsim 0.1$ \citep{Plotnikov2019,Sironi2021},
this tendency holds for 3D. For $\sigma_{tot} \gg 1$, however, the previous studies in pair shocks 
show $f_{\xi} \sim 10^{-3}/\sigma_{tot}$, and thus
the emission efficiency disfavors the SMI in highly magnetized shocks. 
The intense electromagnetic waves excite wakefields (i.e., 
electrostatic plasma waves) via stimulated Raman scattering in the upstream region. Then, the wakefields accelerate the incoming 
electrons and take the kinetic energy from the ion flow. The accelerated electrons provide more energy to the SMI and the emission
efficiency is enhanced. The stronger wakefields are in turn excited, completing the feedback loop. This loop continues until the 
energy equipartition between electrons and ions is achieved in the upstream region. Although electrons transfer only a small fraction
of the incoming kinetic energy, the SMI can indirectly consume the ion kinetic energy and thus $\sigma_{tot}$ controls the emission 
efficiency. This ion-electron coupling can work for highly relativistic shocks $\gamma_1 \gg 1$. For mildly relativistic shocks
(i.e., $\gamma_1$ is the order of unity), 
the amplitude of the wakefield is too small to accelerate the incoming electrons and 
the upstream electron kinetic energy is almost constant. 
Therefore, the ion-electron coupling dose not work and 
the SMI in mildly relativistic shocks becomes less efficient 
\citep{Ligorini2021a,Ligorini2021b}. Since the SMI model generally considers 
highly relativistic shocks, the ion-electron coupling operates and the assumption $f_{\xi} \sim 10^{-3}$ is valid in terms of 
$\gamma_1$ as well.

\begin{figure}[htb]
	\includegraphics[width=8.5cm]{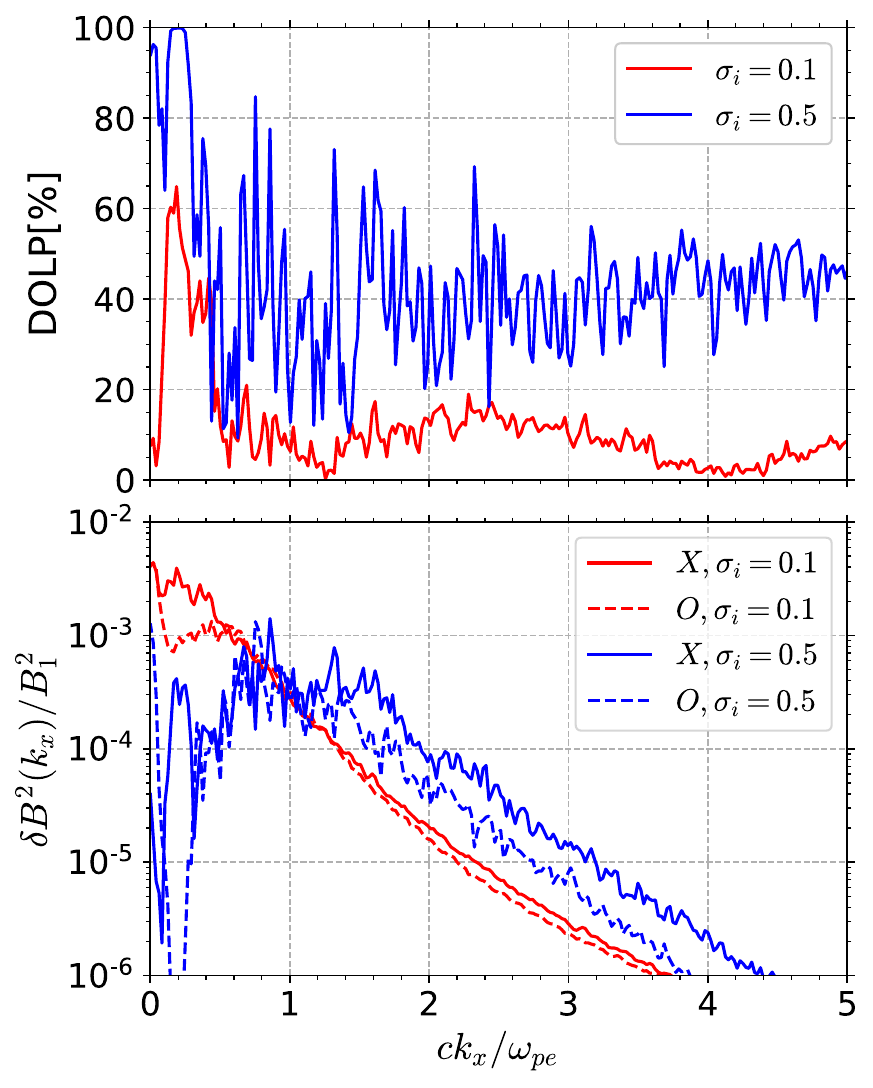}
	\caption{Degree of linear polarization (DOLP) and wave power spectra along the line of sight at the final state $\omega_{pe}t=2000$. 
	The red and blue indicates 
	$\sigma_i=0.1$ and $0.5$, respectively. DOLP (top) is calculated from transversely-averaged Stokes parameters. Power spectra 
	of X (solid lines) and O (dashed lines) mode waves are shown in the bottom panel.}
	\label{fig:fft}
\end{figure}

The degree of linear polarization (DOLP) is shown in the top panel of Figure \ref{fig:fft} for $\sigma_i=0.1$ (red) and $\sigma_i=0.5$ (blue).
We perform Fourier transform of fluctuating magnetic fields along the line of sight ($x$ direction in our coordinates) in the same 
region as $f_\xi$ and calculate the DOLP from the Stokes parameters \citep{Rybicki1979}: I, Q, U and V averaged over the transverse direction. 
The wave power spectra of X (solid lines) and O (dashed lines) mode waves $\bm{\delta B}=\bm{B}-\bm{B_1}$ integrated over the transverse 
wavevector $k_y$ and $k_z$ are shown in the bottom panel. 
The peak of the power spectra at $ck_x/\omega_{pe} \sim 0$ comes from the FI because 
the low-wavenumber waves $ck_x/\omega_{pe} \sim 0$ cannot escape upstream and the fluctuations 
must be induced by the FI in the upstream 
rather than the SMI in the shock transition \citep{Iwamoto2017,Iwamoto2018,Plotnikov2019}. Previous simulations of pair shocks 
\cite{Gallant1992,Iwamoto2017,Iwamoto2018,Plotnikov2019} show the wave power takes the maximum at $ck_x/\omega_{pe} \sim 3-5$, 
which is obviously larger than the ion-electron shocks. This can be explained by the ion-electron coupling. The upstream bulk Lorentz 
factor for electrons becomes $\gamma_{eff} \sim m_i \gamma_1/2m_e$ due to the energy equipartition and the upstream electron plasma 
frequency decreases by a factor of $\sqrt{m_i/2m_e}=10$, resulting in the peak shift \citep{Lyubarsky2021}. The cutoff wavenumber 
below which the electromagnetic waves cannot catch up with the shocks also decreases, and thus a sharp cutoff is not observed unlike 
pair shocks. For $\sigma_i=0.1$, the DOLP is 60\% at $ck_x/\omega_{pe} \sim 0.2$ where the X-mode wave power takes the maximum. 
For $\sigma_i=0.5$, the electromagnetic waves are more highly linearly polarized for the wide range of the wavenumber and the DOLP is 
higher than 60\% at around the peak $ck_x/\omega_{pe} \sim 0.8$. Furthermore, it reaches almost 100\% at $ck_x/\omega_{pe} \sim 0.2$ 
where the electromagnetic waves still have significant power. 
We determine the accurate DOLP of the synchrotron maser emission
based on Stokes parameters and show for the first time that it is considerably high at the dominant modes.

\begin{figure}[htb]
	\includegraphics[width=8.5cm]{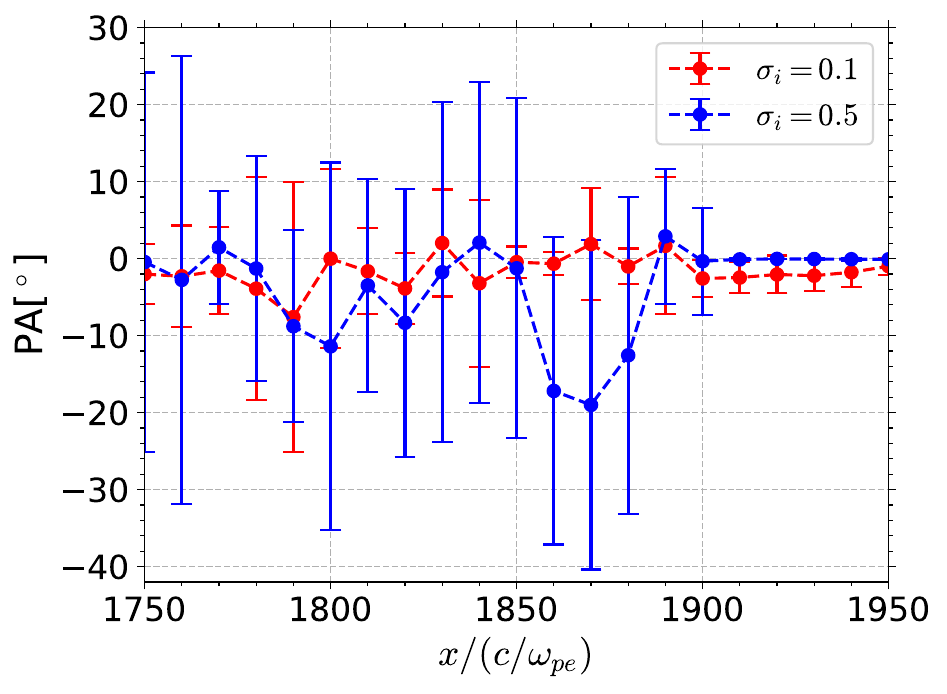}
	\caption{Polarization angle (PA) along the line of sight at the final state $\omega_{pe}t=2000$. 
	The red and blue lines represent $\sigma_i=0.1$ and $0.5$, respectively.
	PA is calculated from transversely-averaged Stokes parameters with the spatial window width $10 c/\omega_{pe}$.}
	\label{fig:pa}
\end{figure}

The polarization angle (PA) of the synchrotron maser emission is no longer constant due to the excitation of the O mode wave. 
The PA in units of degree is determined for the electromagnetic waves that have been emitted at the early phase of the shock evolution: 
$1750 \leq x/(c/\omega_{pe}) \leq 1950$ and shown in Figure \ref{fig:pa} for $\sigma_i=0.1$ (red) and $0.5$ (blue). 
We determine Stokes parameters along the line of sight ($x$ direction)
with the spatial window width $10c/\omega_{pe}$. The PA along the line of sight is calculated from them at each position and then 
averaged over both transverse direction and wavenumber space. The error bars are determined from the standard deviation of the PA. 
For both $\sigma_i$, the PA is almost constant in the region $1900 \leq x/(c/\omega_{pe}) \leq 1950$ because only the X mode waves are 
generated at the early phase of the shock evolution. The O mode waves are induced when the ambient magnetic field is sufficiently 
perturbed by the AIC, and thus the O mode waves lag behind the X mode waves \citep{Iwamoto2019,Ligorini2021a}. Note that the group 
velocities of these two waves are almost equal to the speed of light and the time delay of arrival comes from the difference of 
excitation time. In the region $x/(c/\omega_{pe})\leq 1900$, the PA is strongly modified and the errors become larger due to the 
mixture of the X and O mode waves. Especially for $\sigma_i=0.5$, the PA drastically changes and this change can be observed as the 
PA swing. Since both X and O mode waves are induced in pair shocks as well \citep{Iwamoto2018,Sironi2021}, the 
PA swing can be observed regardless of 
the plasma components. The observations of FRB 180301 indeed show the various PA swings, and this observational fact is believed 
to be an indirect proof that FRBs originate from the coherent curvature emission of electron bunches formed in the magnetar 
magnetosphere \citep{Luo2020}. 
However, our simulations indicate that the PA swings can be reproduced also by the synchrotron maser 
emission and the mixture of the two different linearly polarized waves can result in the diversity of the PA. 

We now discuss the peak frequency of the synchrotron maser emission based on 
the previous work \cite{Plotnikov2019}. 
The peak wavenumber in the downstream rest frame $k_{peak}$ 
can be expressed as (see the bottom panel in Figure \ref{fig:fft})
\begin{equation}
	k_{peak} \sim \zeta \sqrt{\frac{2m_e}{m_i}} \frac{\omega_{pe}}{c},
\end{equation}
where $\zeta$ is a few. The factor $\sqrt{2m_e/m_i}$ comes from the ion-electron coupling \citep{Lyubarsky2021}. Since the dispersion 
relation of the electromagnetic waves in the downstream rest frame is written as $\omega^2=\omega_{pe}^2+c^2 k^2$ for 
$\sigma_e/\gamma_1^2 \gg 1$ \citep{Ligorini2021a}, the peak frequency in the upstream rest frame  $\nu_{peak}$ is 
\begin{equation}
	\nu_{peak} \sim \gamma_1 \sqrt{1+\zeta^2} \sqrt{\frac{2m_e}{m_i}} \frac{\omega_{pe}}{2\pi} 
	\sim 1 {\rm GHz} \sqrt{\frac{\gamma_{sh}^2 n_0}{10^{13} {\rm cm}^{-3}}},
\end{equation}
where $\gamma_{sh}$ and $n_0$ are the shock Lorentz factor and the upstream electron number density measured in the upstream rest frame, 
respectively. Here we have used $\gamma_1 \sim \gamma_{sh}$, which is valid for $\sigma_i \leq 1$, and neglected factors of order of 
unity. If the upstream rest frame corresponds to the observer frame, which is the case for the baryon-loaded shell expanding with the 
non-relativistic speed, $\gamma_{sh}^2 n_0 \sim 10^{13} {\rm cm}^{-3}$ is required for the coherent emission in GHz band. 
The SMI model \citep{Metzger2019,Margalit2019,Margalit2020} can satisfy the condition for a reasonable choice of parameters. 
The obtained emission efficiency $f_\xi$ may change if the upstream plasmas are hot \citep{Babul2020}. Even though the upstream 
plasmas are initially cold, the FI and the stimulated Raman scattering lead to the heating and even to the generation of the 
nonthermal particles in the upstream \citep{Iwamoto2019,Iwamoto2022}. The temperature dependence of the emission efficiency $f_\xi$ 
in ion-electron shocks remains unsolved. Inclusion of positrons may affect the emission efficiency as well. In ion-electron-positron 
plasmas, the resonant interaction between the incoming positrons and ions via the SMI occurs 
in the shock transition and a significant fraction of the ion kinetic energy is preferentially
transferred to the positrons \citep{Hoshino1991,Hoshino1992}. 
Furthermore, wakefields become weaker because both electrons and positrons are pushed by the ponderomotive force, and the ion-electron 
coupling in the upstream region can be inefficient. The emission efficiency in ion-electron-positron shocks is still an open question.

\begin{acknowledgments}

We are grateful to Lorenzo Sironi, Emanuele Sobacchi, Jacek Niemiec, and 
Martin Pohl for fruitful discussions.
This research was supported by MEXT as “Program for Promoting Researches on the Supercomputer 
Fugaku” (Toward a unified view of the universe: from large scale structures to planets, 
JPMXP1020200109) and JICFuS.
MI acknowledges support from JSPS KAKENHI grant No. 20J00280, 20KK0064 and 22H00130.

\end{acknowledgments}


\begin{thebibliography}{47}%
	\makeatletter
	\providecommand \@ifxundefined [1]{%
	 \@ifx{#1\undefined}
	}%
	\providecommand \@ifnum [1]{%
	 \ifnum #1\expandafter \@firstoftwo
	 \else \expandafter \@secondoftwo
	 \fi
	}%
	\providecommand \@ifx [1]{%
	 \ifx #1\expandafter \@firstoftwo
	 \else \expandafter \@secondoftwo
	 \fi
	}%
	\providecommand \natexlab [1]{#1}%
	\providecommand \enquote  [1]{``#1''}%
	\providecommand \bibnamefont  [1]{#1}%
	\providecommand \bibfnamefont [1]{#1}%
	\providecommand \citenamefont [1]{#1}%
	\providecommand \href@noop [0]{\@secondoftwo}%
	\providecommand \href [0]{\begingroup \@sanitize@url \@href}%
	\providecommand \@href[1]{\@@startlink{#1}\@@href}%
	\providecommand \@@href[1]{\endgroup#1\@@endlink}%
	\providecommand \@sanitize@url [0]{\catcode `\\12\catcode `\$12\catcode `\&12\catcode `\#12\catcode `\^12\catcode `\_12\catcode `\%12\relax}%
	\providecommand \@@startlink[1]{}%
	\providecommand \@@endlink[0]{}%
	\providecommand \url  [0]{\begingroup\@sanitize@url \@url }%
	\providecommand \@url [1]{\endgroup\@href {#1}{\urlprefix }}%
	\providecommand \urlprefix  [0]{URL }%
	\providecommand \Eprint [0]{\href }%
	\providecommand \doibase [0]{https://doi.org/}%
	\providecommand \selectlanguage [0]{\@gobble}%
	\providecommand \bibinfo  [0]{\@secondoftwo}%
	\providecommand \bibfield  [0]{\@secondoftwo}%
	\providecommand \translation [1]{[#1]}%
	\providecommand \BibitemOpen [0]{}%
	\providecommand \bibitemStop [0]{}%
	\providecommand \bibitemNoStop [0]{.\EOS\space}%
	\providecommand \EOS [0]{\spacefactor3000\relax}%
	\providecommand \BibitemShut  [1]{\csname bibitem#1\endcsname}%
	\let\auto@bib@innerbib\@empty
	\bibitem [{\citenamefont {Lorimer}\ \emph {et~al.}(2007)\citenamefont {Lorimer}, \citenamefont {Bailes}, \citenamefont {McLaughlin}, \citenamefont {Narkevic},\ and\ \citenamefont {Crawford}}]{Lorimer2007}%
	  \BibitemOpen
	  \bibfield  {author} {\bibinfo {author} {\bibfnamefont {D.~R.}\ \bibnamefont {Lorimer}}, \bibinfo {author} {\bibfnamefont {M.}~\bibnamefont {Bailes}}, \bibinfo {author} {\bibfnamefont {M.~A.}\ \bibnamefont {McLaughlin}}, \bibinfo {author} {\bibfnamefont {D.~J.}\ \bibnamefont {Narkevic}},\ and\ \bibinfo {author} {\bibfnamefont {F.}~\bibnamefont {Crawford}},\ }\bibfield  {title} {\bibinfo {title} {{A Bright Millisecond Radio Burst of Extragalactic Origin}},\ }\href {https://doi.org/10.1126/science.1147532} {\bibfield  {journal} {\bibinfo  {journal} {Science}\ }\textbf {\bibinfo {volume} {318}},\ \bibinfo {pages} {777} (\bibinfo {year} {2007})}\BibitemShut {NoStop}%
	\bibitem [{\citenamefont {Petroff}\ \emph {et~al.}(2022)\citenamefont {Petroff}, \citenamefont {Hessels},\ and\ \citenamefont {Lorimer}}]{Petroff2022}%
	  \BibitemOpen
	  \bibfield  {author} {\bibinfo {author} {\bibfnamefont {E.}~\bibnamefont {Petroff}}, \bibinfo {author} {\bibfnamefont {J.~W.}\ \bibnamefont {Hessels}},\ and\ \bibinfo {author} {\bibfnamefont {D.~R.}\ \bibnamefont {Lorimer}},\ }\bibfield  {title} {\bibinfo {title} {{Fast radio bursts at the dawn of the 2020s}},\ }\href {https://doi.org/10.1007/s00159-022-00139-w} {\bibfield  {journal} {\bibinfo  {journal} {Astron. Astrophys. Rev.}\ }\textbf {\bibinfo {volume} {30}},\ \bibinfo {pages} {1} (\bibinfo {year} {2022})}\BibitemShut {NoStop}%
	\bibitem [{\citenamefont {Katz}(2022{\natexlab{a}})}]{Katz2022b}%
	  \BibitemOpen
	  \bibfield  {author} {\bibinfo {author} {\bibfnamefont {J.~I.}\ \bibnamefont {Katz}},\ }\bibfield  {title} {\bibinfo {title} {{The sources of apparently non-repeating FRB}},\ }\href {https://doi.org/10.1093/mnras/stac2174} {\bibfield  {journal} {\bibinfo  {journal} {Mon. Not. R. Astron. Soc.}\ }\textbf {\bibinfo {volume} {516}},\ \bibinfo {pages} {53} (\bibinfo {year} {2022}{\natexlab{a}})}\BibitemShut {NoStop}%
	\bibitem [{\citenamefont {Bhandari}\ \emph {et~al.}(2023)\citenamefont {Bhandari}, \citenamefont {Gordon}, \citenamefont {Scott}, \citenamefont {Marnoch}, \citenamefont {Sridhar}, \citenamefont {Kumar}, \citenamefont {James}, \citenamefont {Qiu}, \citenamefont {Bannister}, \citenamefont {{T. Deller}}, \citenamefont {Eftekhari}, \citenamefont {Fong}, \citenamefont {Glowacki}, \citenamefont {Prochaska}, \citenamefont {Ryder}, \citenamefont {Shannon},\ and\ \citenamefont {Simha}}]{Bhandari2023}%
	  \BibitemOpen
	  \bibfield  {author} {\bibinfo {author} {\bibfnamefont {S.}~\bibnamefont {Bhandari}}, \bibinfo {author} {\bibfnamefont {A.~C.}\ \bibnamefont {Gordon}}, \bibinfo {author} {\bibfnamefont {D.~R.}\ \bibnamefont {Scott}}, \bibinfo {author} {\bibfnamefont {L.}~\bibnamefont {Marnoch}}, \bibinfo {author} {\bibfnamefont {N.}~\bibnamefont {Sridhar}}, \bibinfo {author} {\bibfnamefont {P.}~\bibnamefont {Kumar}}, \bibinfo {author} {\bibfnamefont {C.~W.}\ \bibnamefont {James}}, \bibinfo {author} {\bibfnamefont {H.}~\bibnamefont {Qiu}}, \bibinfo {author} {\bibfnamefont {K.~W.}\ \bibnamefont {Bannister}}, \bibinfo {author} {\bibfnamefont {A.}~\bibnamefont {{T. Deller}}}, \bibinfo {author} {\bibfnamefont {T.}~\bibnamefont {Eftekhari}}, \bibinfo {author} {\bibfnamefont {W.-f.}\ \bibnamefont {Fong}}, \bibinfo {author} {\bibfnamefont {M.}~\bibnamefont {Glowacki}}, \bibinfo {author} {\bibfnamefont {J.~X.}\ \bibnamefont {Prochaska}}, \bibinfo {author} {\bibfnamefont {S.~D.}\ \bibnamefont {Ryder}}, \bibinfo {author} {\bibfnamefont {R.~M.}\ \bibnamefont {Shannon}},\ and\ \bibinfo {author} {\bibfnamefont {S.}~\bibnamefont {Simha}},\ }\bibfield  {title} {\bibinfo {title} {{A Nonrepeating Fast Radio Burst in a Dwarf Host Galaxy}},\ }\href {https://doi.org/10.3847/1538-4357/acc178} {\bibfield  {journal} {\bibinfo  {journal} {Astrophys. J.}\ }\textbf {\bibinfo {volume} {948}},\ \bibinfo {pages} {67} (\bibinfo {year} {2023})}\BibitemShut {NoStop}%
	\bibitem [{\citenamefont {Lyubarsky}(2021)}]{Lyubarsky2021}%
	  \BibitemOpen
	  \bibfield  {author} {\bibinfo {author} {\bibfnamefont {Y.}~\bibnamefont {Lyubarsky}},\ }\bibfield  {title} {\bibinfo {title} {{Emission Mechanisms of Fast Radio Bursts}},\ }\href {https://doi.org/10.3390/universe7030056} {\bibfield  {journal} {\bibinfo  {journal} {Universe}\ }\textbf {\bibinfo {volume} {7}},\ \bibinfo {pages} {56} (\bibinfo {year} {2021})}\BibitemShut {NoStop}%
	\bibitem [{\citenamefont {Andersen}\ \emph {et~al.}(2020)\citenamefont {Andersen}, \citenamefont {Bandura}, \citenamefont {Bhardwaj}, \citenamefont {Bij}, \citenamefont {Boyce}, \citenamefont {Boyle}, \citenamefont {Brar}, \citenamefont {Cassanelli}, \citenamefont {Chawla}, \citenamefont {Chen}, \citenamefont {Cliche}, \citenamefont {Cook}, \citenamefont {Cubranic}, \citenamefont {Curtin}, \citenamefont {Denman}, \citenamefont {Dobbs}, \citenamefont {Dong}, \citenamefont {Fandino}, \citenamefont {Fonseca}, \citenamefont {Gaensler}, \citenamefont {Giri}, \citenamefont {Good}, \citenamefont {Halpern}, \citenamefont {Hill}, \citenamefont {Hinshaw}, \citenamefont {H{\"{o}}fer}, \citenamefont {Josephy}, \citenamefont {Kania}, \citenamefont {Kaspi}, \citenamefont {Landecker}, \citenamefont {Leung}, \citenamefont {Li}, \citenamefont {Lin}, \citenamefont {Masui}, \citenamefont {Mckinven}, \citenamefont {Mena-Parra}, \citenamefont {Merryfield}, \citenamefont {Meyers}, \citenamefont {Michilli}, \citenamefont {Milutinovic}, \citenamefont {Mirhosseini}, \citenamefont {M{\"{u}}nchmeyer}, \citenamefont {Naidu}, \citenamefont {Newburgh}, \citenamefont {Ng}, \citenamefont {Patel}, \citenamefont {Pen}, \citenamefont {Pinsonneault-Marotte}, \citenamefont {Pleunis}, \citenamefont {Quine}, \citenamefont {Rafiei-Ravandi}, \citenamefont {Rahman}, \citenamefont {Ransom}, \citenamefont {Renard}, \citenamefont {Sanghavi}, \citenamefont {Scholz}, \citenamefont {Shaw}, \citenamefont {Shin}, \citenamefont {Siegel}, \citenamefont {Singh}, \citenamefont {Smegal}, \citenamefont {Smith}, \citenamefont {Stairs}, \citenamefont {Tan}, \citenamefont {Tendulkar}, \citenamefont {Tretyakov}, \citenamefont {Vanderlinde}, \citenamefont {Wang}, \citenamefont {Wulf},\ and\ \citenamefont {Zwaniga}}]{Andersen2020}%
	  \BibitemOpen
	  \bibfield  {author} {\bibinfo {author} {\bibfnamefont {B.~C.}\ \bibnamefont {Andersen}}, \bibinfo {author} {\bibfnamefont {K.~M.}\ \bibnamefont {Bandura}}, \bibinfo {author} {\bibfnamefont {M.}~\bibnamefont {Bhardwaj}}, \bibinfo {author} {\bibfnamefont {A.}~\bibnamefont {Bij}}, \bibinfo {author} {\bibfnamefont {M.~M.}\ \bibnamefont {Boyce}}, \bibinfo {author} {\bibfnamefont {P.~J.}\ \bibnamefont {Boyle}}, \bibinfo {author} {\bibfnamefont {C.}~\bibnamefont {Brar}}, \bibinfo {author} {\bibfnamefont {T.}~\bibnamefont {Cassanelli}}, \bibinfo {author} {\bibfnamefont {P.}~\bibnamefont {Chawla}}, \bibinfo {author} {\bibfnamefont {T.}~\bibnamefont {Chen}}, \bibinfo {author} {\bibfnamefont {J.~F.}\ \bibnamefont {Cliche}}, \bibinfo {author} {\bibfnamefont {A.}~\bibnamefont {Cook}}, \bibinfo {author} {\bibfnamefont {D.}~\bibnamefont {Cubranic}}, \bibinfo {author} {\bibfnamefont {A.~P.}\ \bibnamefont {Curtin}}, \bibinfo {author} {\bibfnamefont {N.~T.}\ \bibnamefont {Denman}}, \bibinfo {author} {\bibfnamefont {M.}~\bibnamefont {Dobbs}}, \bibinfo {author} {\bibfnamefont {F.~Q.}\ \bibnamefont {Dong}}, \bibinfo {author} {\bibfnamefont {M.}~\bibnamefont {Fandino}}, \bibinfo {author} {\bibfnamefont {E.}~\bibnamefont {Fonseca}}, \bibinfo {author} {\bibfnamefont {B.~M.}\ \bibnamefont {Gaensler}}, \bibinfo {author} {\bibfnamefont {U.}~\bibnamefont {Giri}}, \bibinfo {author} {\bibfnamefont {D.~C.}\ \bibnamefont {Good}}, \bibinfo {author} {\bibfnamefont {M.}~\bibnamefont {Halpern}}, \bibinfo {author} {\bibfnamefont {A.~S.}\ \bibnamefont {Hill}}, \bibinfo {author} {\bibfnamefont {G.~F.}\ \bibnamefont {Hinshaw}}, \bibinfo {author} {\bibfnamefont {C.}~\bibnamefont {H{\"{o}}fer}}, \bibinfo {author} {\bibfnamefont {A.}~\bibnamefont {Josephy}}, \bibinfo {author} {\bibfnamefont {J.~W.}\ \bibnamefont {Kania}}, \bibinfo {author} {\bibfnamefont {V.~M.}\ \bibnamefont {Kaspi}}, \bibinfo {author} {\bibfnamefont {T.~L.}\ \bibnamefont {Landecker}}, \bibinfo {author} {\bibfnamefont {C.}~\bibnamefont {Leung}}, \bibinfo {author} {\bibfnamefont {D.~Z.}\ \bibnamefont {Li}}, \bibinfo {author} {\bibfnamefont {H.~H.}\ \bibnamefont {Lin}}, \bibinfo {author} {\bibfnamefont {K.~W.}\ \bibnamefont {Masui}}, \bibinfo {author} {\bibfnamefont {R.}~\bibnamefont {Mckinven}}, \bibinfo {author} {\bibfnamefont {J.}~\bibnamefont {Mena-Parra}}, \bibinfo {author} {\bibfnamefont {M.}~\bibnamefont {Merryfield}}, \bibinfo {author} {\bibfnamefont {B.~W.}\ \bibnamefont {Meyers}}, \bibinfo {author} {\bibfnamefont {D.}~\bibnamefont {Michilli}}, \bibinfo {author} {\bibfnamefont {N.}~\bibnamefont {Milutinovic}}, \bibinfo {author} {\bibfnamefont {A.}~\bibnamefont {Mirhosseini}}, \bibinfo {author} {\bibfnamefont {M.}~\bibnamefont {M{\"{u}}nchmeyer}}, \bibinfo {author} {\bibfnamefont {A.}~\bibnamefont {Naidu}}, \bibinfo {author} {\bibfnamefont {L.~B.}\ \bibnamefont {Newburgh}}, \bibinfo {author} {\bibfnamefont {C.}~\bibnamefont {Ng}}, \bibinfo {author} {\bibfnamefont {C.}~\bibnamefont {Patel}}, \bibinfo {author} {\bibfnamefont {U.~L.}\ \bibnamefont {Pen}}, \bibinfo {author} {\bibfnamefont {T.}~\bibnamefont {Pinsonneault-Marotte}}, \bibinfo {author} {\bibfnamefont {Z.}~\bibnamefont {Pleunis}}, \bibinfo {author} {\bibfnamefont {B.~M.}\ \bibnamefont {Quine}}, \bibinfo {author} {\bibfnamefont {M.}~\bibnamefont {Rafiei-Ravandi}}, \bibinfo {author} {\bibfnamefont {M.}~\bibnamefont {Rahman}}, \bibinfo {author} {\bibfnamefont {S.~M.}\ \bibnamefont {Ransom}}, \bibinfo {author} {\bibfnamefont {A.}~\bibnamefont {Renard}}, \bibinfo {author} {\bibfnamefont {P.}~\bibnamefont {Sanghavi}}, \bibinfo {author} {\bibfnamefont {P.}~\bibnamefont {Scholz}}, \bibinfo {author} {\bibfnamefont {J.~R.}\ \bibnamefont {Shaw}}, \bibinfo {author} {\bibfnamefont {K.}~\bibnamefont {Shin}}, \bibinfo {author} {\bibfnamefont {S.~R.}\ \bibnamefont {Siegel}}, \bibinfo {author} {\bibfnamefont {S.}~\bibnamefont {Singh}}, \bibinfo {author} {\bibfnamefont {R.~J.}\ \bibnamefont {Smegal}}, \bibinfo {author} {\bibfnamefont {K.~M.}\ \bibnamefont {Smith}}, \bibinfo {author} {\bibfnamefont {I.~H.}\ \bibnamefont {Stairs}}, \bibinfo {author} {\bibfnamefont {C.~M.}\ \bibnamefont {Tan}}, \bibinfo {author} {\bibfnamefont {S.~P.}\ \bibnamefont {Tendulkar}}, \bibinfo {author} {\bibfnamefont {I.}~\bibnamefont {Tretyakov}}, \bibinfo {author} {\bibfnamefont {K.}~\bibnamefont {Vanderlinde}}, \bibinfo {author} {\bibfnamefont {H.}~\bibnamefont {Wang}}, \bibinfo {author} {\bibfnamefont {D.}~\bibnamefont {Wulf}},\ and\ \bibinfo {author} {\bibfnamefont {A.~V.}\ \bibnamefont {Zwaniga}},\ }\bibfield  {title} {\bibinfo {title} {{A bright millisecond-duration radio burst from a Galactic magnetar}},\ }\href {https://doi.org/10.1038/s41586-020-2863-y} {\bibfield  {journal} {\bibinfo  {journal} {Nature}\ }\textbf {\bibinfo {volume} {587}},\ \bibinfo {pages} {54} (\bibinfo {year} {2020})}\BibitemShut {NoStop}%
	\bibitem [{\citenamefont {Bochenek}\ \emph {et~al.}(2020)\citenamefont {Bochenek}, \citenamefont {Ravi}, \citenamefont {Belov}, \citenamefont {Hallinan}, \citenamefont {Kocz}, \citenamefont {Kulkarni},\ and\ \citenamefont {McKenna}}]{Bochenek2020}%
	  \BibitemOpen
	  \bibfield  {author} {\bibinfo {author} {\bibfnamefont {C.~D.}\ \bibnamefont {Bochenek}}, \bibinfo {author} {\bibfnamefont {V.}~\bibnamefont {Ravi}}, \bibinfo {author} {\bibfnamefont {K.~V.}\ \bibnamefont {Belov}}, \bibinfo {author} {\bibfnamefont {G.}~\bibnamefont {Hallinan}}, \bibinfo {author} {\bibfnamefont {J.}~\bibnamefont {Kocz}}, \bibinfo {author} {\bibfnamefont {S.~R.}\ \bibnamefont {Kulkarni}},\ and\ \bibinfo {author} {\bibfnamefont {D.~L.}\ \bibnamefont {McKenna}},\ }\bibfield  {title} {\bibinfo {title} {{A fast radio burst associated with a Galactic magnetar}},\ }\href {https://doi.org/10.1038/s41586-020-2872-x} {\bibfield  {journal} {\bibinfo  {journal} {Nature}\ }\textbf {\bibinfo {volume} {587}},\ \bibinfo {pages} {59} (\bibinfo {year} {2020})}\BibitemShut {NoStop}%
	\bibitem [{\citenamefont {Katz}(2014)}]{Katz2014}%
	  \BibitemOpen
	  \bibfield  {author} {\bibinfo {author} {\bibfnamefont {J.~I.}\ \bibnamefont {Katz}},\ }\bibfield  {title} {\bibinfo {title} {{Coherent emission in fast radio bursts}},\ }\href {https://doi.org/10.1103/PhysRevD.89.103009} {\bibfield  {journal} {\bibinfo  {journal} {Phys. Rev. D}\ }\textbf {\bibinfo {volume} {89}},\ \bibinfo {pages} {103009} (\bibinfo {year} {2014})}\BibitemShut {NoStop}%
	\bibitem [{\citenamefont {Lyubarsky}(2014)}]{Lyubarsky2014}%
	  \BibitemOpen
	  \bibfield  {author} {\bibinfo {author} {\bibfnamefont {Y.}~\bibnamefont {Lyubarsky}},\ }\bibfield  {title} {\bibinfo {title} {{A model for fast extragalactic radio bursts}},\ }\href {https://doi.org/10.1093/mnrasl/slu046} {\bibfield  {journal} {\bibinfo  {journal} {Mon. Not. R. Astron. Soc.}\ }\textbf {\bibinfo {volume} {442}},\ \bibinfo {pages} {L9} (\bibinfo {year} {2014})}\BibitemShut {NoStop}%
	\bibitem [{\citenamefont {Beloborodov}(2017)}]{Beloborodov2017}%
	  \BibitemOpen
	  \bibfield  {author} {\bibinfo {author} {\bibfnamefont {A.~M.}\ \bibnamefont {Beloborodov}},\ }\bibfield  {title} {\bibinfo {title} {{A Flaring Magnetar in FRB 121102?}},\ }\href {https://doi.org/10.3847/2041-8213/aa78f3} {\bibfield  {journal} {\bibinfo  {journal} {Astrophys. J.}\ }\textbf {\bibinfo {volume} {843}},\ \bibinfo {pages} {L26} (\bibinfo {year} {2017})}\BibitemShut {NoStop}%
	\bibitem [{\citenamefont {Beloborodov}(2020)}]{Beloborodov2020}%
	  \BibitemOpen
	  \bibfield  {author} {\bibinfo {author} {\bibfnamefont {A.~M.}\ \bibnamefont {Beloborodov}},\ }\bibfield  {title} {\bibinfo {title} {{Blast Waves from Magnetar Flares and Fast Radio Bursts}},\ }\href {https://doi.org/10.3847/1538-4357/ab83eb} {\bibfield  {journal} {\bibinfo  {journal} {Astrophys. J.}\ }\textbf {\bibinfo {volume} {896}},\ \bibinfo {pages} {142} (\bibinfo {year} {2020})}\BibitemShut {NoStop}%
	\bibitem [{\citenamefont {Metzger}\ \emph {et~al.}(2019)\citenamefont {Metzger}, \citenamefont {Margalit},\ and\ \citenamefont {Sironi}}]{Metzger2019}%
	  \BibitemOpen
	  \bibfield  {author} {\bibinfo {author} {\bibfnamefont {B.~D.}\ \bibnamefont {Metzger}}, \bibinfo {author} {\bibfnamefont {B.}~\bibnamefont {Margalit}},\ and\ \bibinfo {author} {\bibfnamefont {L.}~\bibnamefont {Sironi}},\ }\bibfield  {title} {\bibinfo {title} {{Fast radio bursts as synchrotron maser emission from decelerating relativistic blast waves}},\ }\href {https://doi.org/10.1093/mnras/stz700} {\bibfield  {journal} {\bibinfo  {journal} {Mon. Notices Royal Astron. Soc.}\ }\textbf {\bibinfo {volume} {485}},\ \bibinfo {pages} {4091} (\bibinfo {year} {2019})}\BibitemShut {NoStop}%
	\bibitem [{\citenamefont {Margalit}\ \emph {et~al.}(2020{\natexlab{a}})\citenamefont {Margalit}, \citenamefont {Metzger},\ and\ \citenamefont {Sironi}}]{Margalit2019}%
	  \BibitemOpen
	  \bibfield  {author} {\bibinfo {author} {\bibfnamefont {B.}~\bibnamefont {Margalit}}, \bibinfo {author} {\bibfnamefont {B.~D.}\ \bibnamefont {Metzger}},\ and\ \bibinfo {author} {\bibfnamefont {L.}~\bibnamefont {Sironi}},\ }\bibfield  {title} {\bibinfo {title} {{Constraints on the engines of fast radio bursts}},\ }\href {https://doi.org/10.1093/mnras/staa1036} {\bibfield  {journal} {\bibinfo  {journal} {Mon. Not. R. Astron. Soc.}\ }\textbf {\bibinfo {volume} {494}},\ \bibinfo {pages} {4627} (\bibinfo {year} {2020}{\natexlab{a}})}\BibitemShut {NoStop}%
	\bibitem [{\citenamefont {Margalit}\ \emph {et~al.}(2020{\natexlab{b}})\citenamefont {Margalit}, \citenamefont {Beniamini}, \citenamefont {Sridhar},\ and\ \citenamefont {Metzger}}]{Margalit2020}%
	  \BibitemOpen
	  \bibfield  {author} {\bibinfo {author} {\bibfnamefont {B.}~\bibnamefont {Margalit}}, \bibinfo {author} {\bibfnamefont {P.}~\bibnamefont {Beniamini}}, \bibinfo {author} {\bibfnamefont {N.}~\bibnamefont {Sridhar}},\ and\ \bibinfo {author} {\bibfnamefont {B.~D.}\ \bibnamefont {Metzger}},\ }\bibfield  {title} {\bibinfo {title} {{Implications of a "Fast Radio Burst" from a Galactic Magnetar}},\ }\href {https://doi.org/10.3847/2041-8213/abac57} {\bibfield  {journal} {\bibinfo  {journal} {Astrophys. J. Lett.}\ }\textbf {\bibinfo {volume} {899}},\ \bibinfo {pages} {L27} (\bibinfo {year} {2020}{\natexlab{b}})}\BibitemShut {NoStop}%
	\bibitem [{\citenamefont {Langdon}\ \emph {et~al.}(1988)\citenamefont {Langdon}, \citenamefont {Arons},\ and\ \citenamefont {Max}}]{Langdon1988}%
	  \BibitemOpen
	  \bibfield  {author} {\bibinfo {author} {\bibfnamefont {A.~B.}\ \bibnamefont {Langdon}}, \bibinfo {author} {\bibfnamefont {J.}~\bibnamefont {Arons}},\ and\ \bibinfo {author} {\bibfnamefont {C.~E.}\ \bibnamefont {Max}},\ }\bibfield  {title} {\bibinfo {title} {{Structure of Relativistic Magnetosonic Shocks in Electron-Positron Plasmas}},\ }\href {https://doi.org/10.1103/PhysRevLett.61.779} {\bibfield  {journal} {\bibinfo  {journal} {Phys. Rev. Lett.}\ }\textbf {\bibinfo {volume} {61}},\ \bibinfo {pages} {779} (\bibinfo {year} {1988})}\BibitemShut {NoStop}%
	\bibitem [{\citenamefont {Hoshino}\ and\ \citenamefont {Arons}(1991)}]{Hoshino1991}%
	  \BibitemOpen
	  \bibfield  {author} {\bibinfo {author} {\bibfnamefont {M.}~\bibnamefont {Hoshino}}\ and\ \bibinfo {author} {\bibfnamefont {J.}~\bibnamefont {Arons}},\ }\bibfield  {title} {\bibinfo {title} {{Preferential positron heating and acceleration by synchrotron maser instabilities in relativistic positron--electron--proton plasmas}},\ }\href {https://doi.org/10.1063/1.859877} {\bibfield  {journal} {\bibinfo  {journal} {Phys. Fluids B}\ }\textbf {\bibinfo {volume} {3}},\ \bibinfo {pages} {818} (\bibinfo {year} {1991})}\BibitemShut {NoStop}%
	\bibitem [{\citenamefont {Hoshino}\ \emph {et~al.}(1992)\citenamefont {Hoshino}, \citenamefont {Arons}, \citenamefont {Gallant},\ and\ \citenamefont {Langdon}}]{Hoshino1992}%
	  \BibitemOpen
	  \bibfield  {author} {\bibinfo {author} {\bibfnamefont {M.}~\bibnamefont {Hoshino}}, \bibinfo {author} {\bibfnamefont {J.}~\bibnamefont {Arons}}, \bibinfo {author} {\bibfnamefont {Y.~A.}\ \bibnamefont {Gallant}},\ and\ \bibinfo {author} {\bibfnamefont {A.~B.}\ \bibnamefont {Langdon}},\ }\bibfield  {title} {\bibinfo {title} {{Relativistic magnetosonic shock waves in synchrotron sources - Shock structure and nonthermal acceleration of positrons}},\ }\href {https://doi.org/10.1086/171296} {\bibfield  {journal} {\bibinfo  {journal} {Astrophys. J.}\ }\textbf {\bibinfo {volume} {390}},\ \bibinfo {pages} {454} (\bibinfo {year} {1992})}\BibitemShut {NoStop}%
	\bibitem [{\citenamefont {Gallant}\ \emph {et~al.}(1992)\citenamefont {Gallant}, \citenamefont {Hoshino}, \citenamefont {Langdon}, \citenamefont {Arons},\ and\ \citenamefont {Max}}]{Gallant1992}%
	  \BibitemOpen
	  \bibfield  {author} {\bibinfo {author} {\bibfnamefont {Y.~A.}\ \bibnamefont {Gallant}}, \bibinfo {author} {\bibfnamefont {M.}~\bibnamefont {Hoshino}}, \bibinfo {author} {\bibfnamefont {A.~B.}\ \bibnamefont {Langdon}}, \bibinfo {author} {\bibfnamefont {J.}~\bibnamefont {Arons}},\ and\ \bibinfo {author} {\bibfnamefont {C.~E.}\ \bibnamefont {Max}},\ }\bibfield  {title} {\bibinfo {title} {Relativistic, perpendicular shocks in electron-positron plasmas},\ }\href {https://doi.org/10.1086/171326} {\bibfield  {journal} {\bibinfo  {journal} {Astrophys. J.}\ }\textbf {\bibinfo {volume} {391}},\ \bibinfo {pages} {73} (\bibinfo {year} {1992})}\BibitemShut {NoStop}%
	\bibitem [{\citenamefont {Amato}\ and\ \citenamefont {Arons}(2006)}]{Amato2006}%
	  \BibitemOpen
	  \bibfield  {author} {\bibinfo {author} {\bibfnamefont {E.}~\bibnamefont {Amato}}\ and\ \bibinfo {author} {\bibfnamefont {J.}~\bibnamefont {Arons}},\ }\bibfield  {title} {\bibinfo {title} {Heating and nonthermal particle acceleration in relativistic, transverse magnetosonic shock waves in proton-electron-positron plasmas},\ }\href {https://doi.org/10.1086/508050} {\bibfield  {journal} {\bibinfo  {journal} {Astrophys. J.}\ }\textbf {\bibinfo {volume} {653}},\ \bibinfo {pages} {325} (\bibinfo {year} {2006})}\BibitemShut {NoStop}%
	\bibitem [{\citenamefont {Sironi}\ and\ \citenamefont {Spitkovsky}(2011)}]{Sironi2011}%
	  \BibitemOpen
	  \bibfield  {author} {\bibinfo {author} {\bibfnamefont {L.}~\bibnamefont {Sironi}}\ and\ \bibinfo {author} {\bibfnamefont {A.}~\bibnamefont {Spitkovsky}},\ }\bibfield  {title} {\bibinfo {title} {{PARTICLE ACCELERATION IN RELATIVISTIC MAGNETIZED COLLISIONLESS ELECTRON-ION SHOCKS}},\ }\href {https://doi.org/10.1088/0004-637X/726/2/75} {\bibfield  {journal} {\bibinfo  {journal} {Astrophys. J.}\ }\textbf {\bibinfo {volume} {726}},\ \bibinfo {pages} {75} (\bibinfo {year} {2011})}\BibitemShut {NoStop}%
	\bibitem [{\citenamefont {Iwamoto}\ \emph {et~al.}(2017)\citenamefont {Iwamoto}, \citenamefont {Amano}, \citenamefont {Hoshino},\ and\ \citenamefont {Matsumoto}}]{Iwamoto2017}%
	  \BibitemOpen
	  \bibfield  {author} {\bibinfo {author} {\bibfnamefont {M.}~\bibnamefont {Iwamoto}}, \bibinfo {author} {\bibfnamefont {T.}~\bibnamefont {Amano}}, \bibinfo {author} {\bibfnamefont {M.}~\bibnamefont {Hoshino}},\ and\ \bibinfo {author} {\bibfnamefont {Y.}~\bibnamefont {Matsumoto}},\ }\bibfield  {title} {\bibinfo {title} {{Persistence of Precursor Waves in Two-dimensional Relativistic Shocks}},\ }\href {https://doi.org/10.3847/1538-4357/aa6d6f} {\bibfield  {journal} {\bibinfo  {journal} {Astrophys. J.}\ }\textbf {\bibinfo {volume} {840}},\ \bibinfo {pages} {52} (\bibinfo {year} {2017})}\BibitemShut {NoStop}%
	\bibitem [{\citenamefont {Iwamoto}\ \emph {et~al.}(2018)\citenamefont {Iwamoto}, \citenamefont {Amano}, \citenamefont {Hoshino},\ and\ \citenamefont {Matsumoto}}]{Iwamoto2018}%
	  \BibitemOpen
	  \bibfield  {author} {\bibinfo {author} {\bibfnamefont {M.}~\bibnamefont {Iwamoto}}, \bibinfo {author} {\bibfnamefont {T.}~\bibnamefont {Amano}}, \bibinfo {author} {\bibfnamefont {M.}~\bibnamefont {Hoshino}},\ and\ \bibinfo {author} {\bibfnamefont {Y.}~\bibnamefont {Matsumoto}},\ }\bibfield  {title} {\bibinfo {title} {{Precursor Wave Emission Enhanced by Weibel Instability in Relativistic Shocks}},\ }\href {https://doi.org/10.3847/1538-4357/aaba7a} {\bibfield  {journal} {\bibinfo  {journal} {Astrophys. J.}\ }\textbf {\bibinfo {volume} {858}},\ \bibinfo {pages} {93} (\bibinfo {year} {2018})}\BibitemShut {NoStop}%
	\bibitem [{\citenamefont {Plotnikov}\ and\ \citenamefont {Sironi}(2019)}]{Plotnikov2019}%
	  \BibitemOpen
	  \bibfield  {author} {\bibinfo {author} {\bibfnamefont {I.}~\bibnamefont {Plotnikov}}\ and\ \bibinfo {author} {\bibfnamefont {L.}~\bibnamefont {Sironi}},\ }\bibfield  {title} {\bibinfo {title} {{The synchrotron maser emission from relativistic shocks in Fast Radio Bursts: 1D PIC simulations of cold pair plasmas}},\ }\href {https://doi.org/10.1093/mnras/stz640} {\bibfield  {journal} {\bibinfo  {journal} {Mon. Not. R. Astron. Soc.}\ }\textbf {\bibinfo {volume} {485}},\ \bibinfo {pages} {3816} (\bibinfo {year} {2019})}\BibitemShut {NoStop}%
	\bibitem [{\citenamefont {Sironi}\ \emph {et~al.}(2021)\citenamefont {Sironi}, \citenamefont {Plotnikov}, \citenamefont {N{\"{a}}ttil{\"{a}}},\ and\ \citenamefont {Beloborodov}}]{Sironi2021}%
	  \BibitemOpen
	  \bibfield  {author} {\bibinfo {author} {\bibfnamefont {L.}~\bibnamefont {Sironi}}, \bibinfo {author} {\bibfnamefont {I.}~\bibnamefont {Plotnikov}}, \bibinfo {author} {\bibfnamefont {J.}~\bibnamefont {N{\"{a}}ttil{\"{a}}}},\ and\ \bibinfo {author} {\bibfnamefont {A.~M.}\ \bibnamefont {Beloborodov}},\ }\bibfield  {title} {\bibinfo {title} {{Coherent Electromagnetic Emission from Relativistic Magnetized Shocks}},\ }\href {https://doi.org/10.1103/PhysRevLett.127.035101} {\bibfield  {journal} {\bibinfo  {journal} {Phys. Rev. Lett.}\ }\textbf {\bibinfo {volume} {127}},\ \bibinfo {pages} {035101} (\bibinfo {year} {2021})}\BibitemShut {NoStop}%
	\bibitem [{\citenamefont {Mckinven}\ \emph {et~al.}(2023)\citenamefont {Mckinven}, \citenamefont {Gaensler}, \citenamefont {Michilli}, \citenamefont {Masui}, \citenamefont {Kaspi}, \citenamefont {Su}, \citenamefont {Bhardwaj}, \citenamefont {Cassanelli}, \citenamefont {Chawla}, \citenamefont {Dong}, \citenamefont {Fonseca}, \citenamefont {Leung}, \citenamefont {Li}, \citenamefont {Ng}, \citenamefont {Patel}, \citenamefont {Pearlman}, \citenamefont {Petroff}, \citenamefont {Pleunis}, \citenamefont {Rafiei-Ravandi}, \citenamefont {Rahman}, \citenamefont {Sand}, \citenamefont {Shin}, \citenamefont {Stairs},\ and\ \citenamefont {Tendulkar}}]{Mckinven2023}%
	  \BibitemOpen
	  \bibfield  {author} {\bibinfo {author} {\bibfnamefont {R.}~\bibnamefont {Mckinven}}, \bibinfo {author} {\bibfnamefont {B.~M.}\ \bibnamefont {Gaensler}}, \bibinfo {author} {\bibfnamefont {D.}~\bibnamefont {Michilli}}, \bibinfo {author} {\bibfnamefont {K.}~\bibnamefont {Masui}}, \bibinfo {author} {\bibfnamefont {V.~M.}\ \bibnamefont {Kaspi}}, \bibinfo {author} {\bibfnamefont {J.}~\bibnamefont {Su}}, \bibinfo {author} {\bibfnamefont {M.}~\bibnamefont {Bhardwaj}}, \bibinfo {author} {\bibfnamefont {T.}~\bibnamefont {Cassanelli}}, \bibinfo {author} {\bibfnamefont {P.}~\bibnamefont {Chawla}}, \bibinfo {author} {\bibfnamefont {F.~A.}\ \bibnamefont {Dong}}, \bibinfo {author} {\bibfnamefont {E.}~\bibnamefont {Fonseca}}, \bibinfo {author} {\bibfnamefont {C.}~\bibnamefont {Leung}}, \bibinfo {author} {\bibfnamefont {D.~Z.}\ \bibnamefont {Li}}, \bibinfo {author} {\bibfnamefont {C.}~\bibnamefont {Ng}}, \bibinfo {author} {\bibfnamefont {C.}~\bibnamefont {Patel}}, \bibinfo {author} {\bibfnamefont {A.~B.}\ \bibnamefont {Pearlman}}, \bibinfo {author} {\bibfnamefont {E.}~\bibnamefont {Petroff}}, \bibinfo {author} {\bibfnamefont {Z.}~\bibnamefont {Pleunis}}, \bibinfo {author} {\bibfnamefont {M.}~\bibnamefont {Rafiei-Ravandi}}, \bibinfo {author} {\bibfnamefont {M.}~\bibnamefont {Rahman}}, \bibinfo {author} {\bibfnamefont {K.~R.}\ \bibnamefont {Sand}}, \bibinfo {author} {\bibfnamefont {K.}~\bibnamefont {Shin}}, \bibinfo {author} {\bibfnamefont {I.~H.}\ \bibnamefont {Stairs}},\ and\ \bibinfo {author} {\bibfnamefont {S.}~\bibnamefont {Tendulkar}},\ }\bibfield  {title} {\bibinfo {title} {{Revealing the Dynamic Magnetoionic Environments of Repeating Fast Radio Burst Sources through Multiyear Polarimetric Monitoring with CHIME/FRB}},\ }\href {https://doi.org/10.3847/1538-4357/acd188} {\bibfield  {journal} {\bibinfo  {journal} {Astrophys. J.}\ }\textbf {\bibinfo {volume} {951}},\ \bibinfo {pages} {82} (\bibinfo {year} {2023})}\BibitemShut {NoStop}%
	\bibitem [{\citenamefont {Masui}\ \emph {et~al.}(2015)\citenamefont {Masui}, \citenamefont {Lin}, \citenamefont {Sievers}, \citenamefont {Anderson}, \citenamefont {Chang}, \citenamefont {Chen}, \citenamefont {Ganguly}, \citenamefont {Jarvis}, \citenamefont {Kuo}, \citenamefont {Li}, \citenamefont {Liao}, \citenamefont {McLaughlin}, \citenamefont {Pen}, \citenamefont {Peterson}, \citenamefont {Roman}, \citenamefont {Timbie}, \citenamefont {Voytek},\ and\ \citenamefont {Yadav}}]{Masui2015}%
	  \BibitemOpen
	  \bibfield  {author} {\bibinfo {author} {\bibfnamefont {K.}~\bibnamefont {Masui}}, \bibinfo {author} {\bibfnamefont {H.~H.}\ \bibnamefont {Lin}}, \bibinfo {author} {\bibfnamefont {J.}~\bibnamefont {Sievers}}, \bibinfo {author} {\bibfnamefont {C.~J.}\ \bibnamefont {Anderson}}, \bibinfo {author} {\bibfnamefont {T.~C.}\ \bibnamefont {Chang}}, \bibinfo {author} {\bibfnamefont {X.}~\bibnamefont {Chen}}, \bibinfo {author} {\bibfnamefont {A.}~\bibnamefont {Ganguly}}, \bibinfo {author} {\bibfnamefont {M.}~\bibnamefont {Jarvis}}, \bibinfo {author} {\bibfnamefont {C.~Y.}\ \bibnamefont {Kuo}}, \bibinfo {author} {\bibfnamefont {Y.~C.}\ \bibnamefont {Li}}, \bibinfo {author} {\bibfnamefont {Y.~W.}\ \bibnamefont {Liao}}, \bibinfo {author} {\bibfnamefont {M.}~\bibnamefont {McLaughlin}}, \bibinfo {author} {\bibfnamefont {U.~L.}\ \bibnamefont {Pen}}, \bibinfo {author} {\bibfnamefont {J.~B.}\ \bibnamefont {Peterson}}, \bibinfo {author} {\bibfnamefont {A.}~\bibnamefont {Roman}}, \bibinfo {author} {\bibfnamefont {P.~T.}\ \bibnamefont {Timbie}}, \bibinfo {author} {\bibfnamefont {T.}~\bibnamefont {Voytek}},\ and\ \bibinfo {author} {\bibfnamefont {J.~K.}\ \bibnamefont {Yadav}},\ }\bibfield  {title} {\bibinfo {title} {{Dense magnetized plasma associated with a fast radio burst}},\ }\href {https://doi.org/10.1038/nature15769} {\bibfield  {journal} {\bibinfo  {journal} {Nature}\ }\textbf {\bibinfo {volume} {528}},\ \bibinfo {pages} {523} (\bibinfo {year} {2015})}\BibitemShut {NoStop}%
	\bibitem [{\citenamefont {Michilli}\ \emph {et~al.}(2018)\citenamefont {Michilli}, \citenamefont {Seymour}, \citenamefont {Hessels}, \citenamefont {Spitler}, \citenamefont {Gajjar}, \citenamefont {Archibald}, \citenamefont {Bower}, \citenamefont {Chatterjee}, \citenamefont {Cordes}, \citenamefont {Gourdji}, \citenamefont {Heald}, \citenamefont {Kaspi}, \citenamefont {Law}, \citenamefont {Sobey}, \citenamefont {Adams}, \citenamefont {Bassa}, \citenamefont {Bogdanov}, \citenamefont {Brinkman}, \citenamefont {Demorest}, \citenamefont {Fernandez}, \citenamefont {Hellbourg}, \citenamefont {Lazio}, \citenamefont {Lynch}, \citenamefont {Maddox}, \citenamefont {Marcote}, \citenamefont {McLaughlin}, \citenamefont {Paragi}, \citenamefont {Ransom}, \citenamefont {Scholz}, \citenamefont {Siemion}, \citenamefont {Tendulkar}, \citenamefont {{Van Rooy}}, \citenamefont {Wharton},\ and\ \citenamefont {Whitlow}}]{Michilli2018}%
	  \BibitemOpen
	  \bibfield  {author} {\bibinfo {author} {\bibfnamefont {D.}~\bibnamefont {Michilli}}, \bibinfo {author} {\bibfnamefont {A.}~\bibnamefont {Seymour}}, \bibinfo {author} {\bibfnamefont {J.~W.}\ \bibnamefont {Hessels}}, \bibinfo {author} {\bibfnamefont {L.~G.}\ \bibnamefont {Spitler}}, \bibinfo {author} {\bibfnamefont {V.}~\bibnamefont {Gajjar}}, \bibinfo {author} {\bibfnamefont {A.~M.}\ \bibnamefont {Archibald}}, \bibinfo {author} {\bibfnamefont {G.~C.}\ \bibnamefont {Bower}}, \bibinfo {author} {\bibfnamefont {S.}~\bibnamefont {Chatterjee}}, \bibinfo {author} {\bibfnamefont {J.~M.}\ \bibnamefont {Cordes}}, \bibinfo {author} {\bibfnamefont {K.}~\bibnamefont {Gourdji}}, \bibinfo {author} {\bibfnamefont {G.~H.}\ \bibnamefont {Heald}}, \bibinfo {author} {\bibfnamefont {V.~M.}\ \bibnamefont {Kaspi}}, \bibinfo {author} {\bibfnamefont {C.~J.}\ \bibnamefont {Law}}, \bibinfo {author} {\bibfnamefont {C.}~\bibnamefont {Sobey}}, \bibinfo {author} {\bibfnamefont {E.~A.}\ \bibnamefont {Adams}}, \bibinfo {author} {\bibfnamefont {C.~G.}\ \bibnamefont {Bassa}}, \bibinfo {author} {\bibfnamefont {S.}~\bibnamefont {Bogdanov}}, \bibinfo {author} {\bibfnamefont {C.}~\bibnamefont {Brinkman}}, \bibinfo {author} {\bibfnamefont {P.}~\bibnamefont {Demorest}}, \bibinfo {author} {\bibfnamefont {F.}~\bibnamefont {Fernandez}}, \bibinfo {author} {\bibfnamefont {G.}~\bibnamefont {Hellbourg}}, \bibinfo {author} {\bibfnamefont {T.~J.}\ \bibnamefont {Lazio}}, \bibinfo {author} {\bibfnamefont {R.~S.}\ \bibnamefont {Lynch}}, \bibinfo {author} {\bibfnamefont {N.}~\bibnamefont {Maddox}}, \bibinfo {author} {\bibfnamefont {B.}~\bibnamefont {Marcote}}, \bibinfo {author} {\bibfnamefont {M.~A.}\ \bibnamefont {McLaughlin}}, \bibinfo {author} {\bibfnamefont {Z.}~\bibnamefont {Paragi}}, \bibinfo {author} {\bibfnamefont {S.~M.}\ \bibnamefont {Ransom}}, \bibinfo {author} {\bibfnamefont {P.}~\bibnamefont {Scholz}}, \bibinfo {author} {\bibfnamefont {A.~P.}\ \bibnamefont {Siemion}}, \bibinfo {author} {\bibfnamefont {S.~P.}\ \bibnamefont {Tendulkar}}, \bibinfo {author} {\bibfnamefont {P.}~\bibnamefont {{Van Rooy}}}, \bibinfo {author} {\bibfnamefont {R.~S.}\ \bibnamefont {Wharton}},\ and\ \bibinfo {author} {\bibfnamefont {D.}~\bibnamefont {Whitlow}},\ }\bibfield  {title} {\bibinfo {title} {{An extreme magneto-ionic environment associated with the fast radio burst source FRB 121102}},\ }\href {https://doi.org/10.1038/nature25149} {\bibfield  {journal} {\bibinfo  {journal} {Nature}\ }\textbf {\bibinfo {volume} {553}},\ \bibinfo {pages} {182} (\bibinfo {year} {2018})}\BibitemShut {NoStop}%
	\bibitem [{\citenamefont {Petroff}\ \emph {et~al.}(2019)\citenamefont {Petroff}, \citenamefont {Hessels},\ and\ \citenamefont {Lorimer}}]{Petroff2019}%
	  \BibitemOpen
	  \bibfield  {author} {\bibinfo {author} {\bibfnamefont {E.}~\bibnamefont {Petroff}}, \bibinfo {author} {\bibfnamefont {J.~W.~T.}\ \bibnamefont {Hessels}},\ and\ \bibinfo {author} {\bibfnamefont {D.~R.}\ \bibnamefont {Lorimer}},\ }\bibfield  {title} {\bibinfo {title} {{Fast radio bursts}},\ }\href {https://doi.org/10.1007/s00159-019-0116-6} {\bibfield  {journal} {\bibinfo  {journal} {Astron. Astrophys. Rev.}\ }\textbf {\bibinfo {volume} {27}},\ \bibinfo {pages} {4} (\bibinfo {year} {2019})}\BibitemShut {NoStop}%
	\bibitem [{\citenamefont {Luo}\ \emph {et~al.}(2020)\citenamefont {Luo}, \citenamefont {Wang}, \citenamefont {Men}, \citenamefont {Zhang}, \citenamefont {Jiang}, \citenamefont {Xu}, \citenamefont {Wang}, \citenamefont {Lee}, \citenamefont {Han}, \citenamefont {Zhang}, \citenamefont {Caballero}, \citenamefont {Chen}, \citenamefont {Chen}, \citenamefont {Gan}, \citenamefont {Guo}, \citenamefont {Hao}, \citenamefont {Huang}, \citenamefont {Jiang}, \citenamefont {Li}, \citenamefont {Li}, \citenamefont {Li}, \citenamefont {Luo}, \citenamefont {Pan}, \citenamefont {Pei}, \citenamefont {Qian}, \citenamefont {Sun}, \citenamefont {Wang}, \citenamefont {Wang}, \citenamefont {Wen}, \citenamefont {Xu}, \citenamefont {Xu}, \citenamefont {Yan}, \citenamefont {Yan}, \citenamefont {Yu}, \citenamefont {Yuan}, \citenamefont {Zhang},\ and\ \citenamefont {Zhu}}]{Luo2020}%
	  \BibitemOpen
	  \bibfield  {author} {\bibinfo {author} {\bibfnamefont {R.}~\bibnamefont {Luo}}, \bibinfo {author} {\bibfnamefont {B.~J.}\ \bibnamefont {Wang}}, \bibinfo {author} {\bibfnamefont {Y.~P.}\ \bibnamefont {Men}}, \bibinfo {author} {\bibfnamefont {C.~F.}\ \bibnamefont {Zhang}}, \bibinfo {author} {\bibfnamefont {J.~C.}\ \bibnamefont {Jiang}}, \bibinfo {author} {\bibfnamefont {H.}~\bibnamefont {Xu}}, \bibinfo {author} {\bibfnamefont {W.~Y.}\ \bibnamefont {Wang}}, \bibinfo {author} {\bibfnamefont {K.~J.}\ \bibnamefont {Lee}}, \bibinfo {author} {\bibfnamefont {J.~L.}\ \bibnamefont {Han}}, \bibinfo {author} {\bibfnamefont {B.}~\bibnamefont {Zhang}}, \bibinfo {author} {\bibfnamefont {R.~N.}\ \bibnamefont {Caballero}}, \bibinfo {author} {\bibfnamefont {M.~Z.}\ \bibnamefont {Chen}}, \bibinfo {author} {\bibfnamefont {X.~L.}\ \bibnamefont {Chen}}, \bibinfo {author} {\bibfnamefont {H.~Q.}\ \bibnamefont {Gan}}, \bibinfo {author} {\bibfnamefont {Y.~J.}\ \bibnamefont {Guo}}, \bibinfo {author} {\bibfnamefont {L.~F.}\ \bibnamefont {Hao}}, \bibinfo {author} {\bibfnamefont {Y.~X.}\ \bibnamefont {Huang}}, \bibinfo {author} {\bibfnamefont {P.}~\bibnamefont {Jiang}}, \bibinfo {author} {\bibfnamefont {H.}~\bibnamefont {Li}}, \bibinfo {author} {\bibfnamefont {J.}~\bibnamefont {Li}}, \bibinfo {author} {\bibfnamefont {Z.~X.}\ \bibnamefont {Li}}, \bibinfo {author} {\bibfnamefont {J.~T.}\ \bibnamefont {Luo}}, \bibinfo {author} {\bibfnamefont {J.}~\bibnamefont {Pan}}, \bibinfo {author} {\bibfnamefont {X.}~\bibnamefont {Pei}}, \bibinfo {author} {\bibfnamefont {L.}~\bibnamefont {Qian}}, \bibinfo {author} {\bibfnamefont {J.~H.}\ \bibnamefont {Sun}}, \bibinfo {author} {\bibfnamefont {M.}~\bibnamefont {Wang}}, \bibinfo {author} {\bibfnamefont {N.}~\bibnamefont {Wang}}, \bibinfo {author} {\bibfnamefont {Z.~G.}\ \bibnamefont {Wen}}, \bibinfo {author} {\bibfnamefont {R.~X.}\ \bibnamefont {Xu}}, \bibinfo {author} {\bibfnamefont {Y.~H.}\ \bibnamefont {Xu}}, \bibinfo {author} {\bibfnamefont {J.}~\bibnamefont {Yan}}, \bibinfo {author} {\bibfnamefont {W.~M.}\ \bibnamefont {Yan}}, \bibinfo {author} {\bibfnamefont {D.~J.}\ \bibnamefont {Yu}}, \bibinfo {author} {\bibfnamefont {J.~P.}\ \bibnamefont {Yuan}}, \bibinfo {author} {\bibfnamefont {S.~B.}\ \bibnamefont {Zhang}},\ and\ \bibinfo {author} {\bibfnamefont {Y.}~\bibnamefont {Zhu}},\ }\bibfield  {title} {\bibinfo {title} {{Diverse polarization angle swings from a repeating fast radio burst source}},\ }\href {https://doi.org/10.1038/s41586-020-2827-2} {\bibfield  {journal} {\bibinfo  {journal} {Nature}\ }\textbf {\bibinfo {volume} {586}},\ \bibinfo {pages} {693} (\bibinfo {year} {2020})}\BibitemShut {NoStop}%
	\bibitem [{\citenamefont {Day}\ \emph {et~al.}(2020)\citenamefont {Day}, \citenamefont {Deller}, \citenamefont {Shannon}, \citenamefont {Qiu}, \citenamefont {Bannister}, \citenamefont {Bhandari}, \citenamefont {Ekers}, \citenamefont {Flynn}, \citenamefont {James}, \citenamefont {Macquart}, \citenamefont {Mahony}, \citenamefont {Phillips},\ and\ \citenamefont {{Xavier Prochaska}}}]{Day2020}%
	  \BibitemOpen
	  \bibfield  {author} {\bibinfo {author} {\bibfnamefont {C.~K.}\ \bibnamefont {Day}}, \bibinfo {author} {\bibfnamefont {A.~T.}\ \bibnamefont {Deller}}, \bibinfo {author} {\bibfnamefont {R.~M.}\ \bibnamefont {Shannon}}, \bibinfo {author} {\bibfnamefont {H.}~\bibnamefont {Qiu}}, \bibinfo {author} {\bibfnamefont {K.~W.}\ \bibnamefont {Bannister}}, \bibinfo {author} {\bibfnamefont {S.}~\bibnamefont {Bhandari}}, \bibinfo {author} {\bibfnamefont {R.}~\bibnamefont {Ekers}}, \bibinfo {author} {\bibfnamefont {C.}~\bibnamefont {Flynn}}, \bibinfo {author} {\bibfnamefont {C.~W.}\ \bibnamefont {James}}, \bibinfo {author} {\bibfnamefont {J.-P.}\ \bibnamefont {Macquart}}, \bibinfo {author} {\bibfnamefont {E.~K.}\ \bibnamefont {Mahony}}, \bibinfo {author} {\bibfnamefont {C.~J.}\ \bibnamefont {Phillips}},\ and\ \bibinfo {author} {\bibfnamefont {J.}~\bibnamefont {{Xavier Prochaska}}},\ }\bibfield  {title} {\bibinfo {title} {{High time resolution and polarization properties of ASKAP-localized fast radio bursts}},\ }\href {https://doi.org/10.1093/mnras/staa2138} {\bibfield  {journal} {\bibinfo  {journal} {Mon. Not. R. Astron. Soc.}\ }\textbf {\bibinfo {volume} {497}},\ \bibinfo {pages} {3335} (\bibinfo {year} {2020})}\BibitemShut {NoStop}%
	\bibitem [{\citenamefont {Ligorini}\ \emph {et~al.}(2021{\natexlab{a}})\citenamefont {Ligorini}, \citenamefont {Niemiec}, \citenamefont {Kobzar}, \citenamefont {Iwamoto}, \citenamefont {Bohdan}, \citenamefont {Pohl}, \citenamefont {Matsumoto}, \citenamefont {Amano}, \citenamefont {Matsukiyo}, \citenamefont {Esaki},\ and\ \citenamefont {Hoshino}}]{Ligorini2021a}%
	  \BibitemOpen
	  \bibfield  {author} {\bibinfo {author} {\bibfnamefont {A.}~\bibnamefont {Ligorini}}, \bibinfo {author} {\bibfnamefont {J.}~\bibnamefont {Niemiec}}, \bibinfo {author} {\bibfnamefont {O.}~\bibnamefont {Kobzar}}, \bibinfo {author} {\bibfnamefont {M.}~\bibnamefont {Iwamoto}}, \bibinfo {author} {\bibfnamefont {A.}~\bibnamefont {Bohdan}}, \bibinfo {author} {\bibfnamefont {M.}~\bibnamefont {Pohl}}, \bibinfo {author} {\bibfnamefont {Y.}~\bibnamefont {Matsumoto}}, \bibinfo {author} {\bibfnamefont {T.}~\bibnamefont {Amano}}, \bibinfo {author} {\bibfnamefont {S.}~\bibnamefont {Matsukiyo}}, \bibinfo {author} {\bibfnamefont {Y.}~\bibnamefont {Esaki}},\ and\ \bibinfo {author} {\bibfnamefont {M.}~\bibnamefont {Hoshino}},\ }\bibfield  {title} {\bibinfo {title} {{Mildly relativistic magnetized shocks in electron--ion plasmas -- I. Electromagnetic shock structure}},\ }\href {https://doi.org/10.1093/mnras/staa3901} {\bibfield  {journal} {\bibinfo  {journal} {Mon. Not. R. Astron. Soc.}\ }\textbf {\bibinfo {volume} {501}},\ \bibinfo {pages} {4837} (\bibinfo {year} {2021}{\natexlab{a}})}\BibitemShut {NoStop}%
	\bibitem [{\citenamefont {Ligorini}\ \emph {et~al.}(2021{\natexlab{b}})\citenamefont {Ligorini}, \citenamefont {Niemiec}, \citenamefont {Kobzar}, \citenamefont {Iwamoto}, \citenamefont {Bohdan}, \citenamefont {Pohl}, \citenamefont {Matsumoto}, \citenamefont {Amano}, \citenamefont {Matsukiyo},\ and\ \citenamefont {Hoshino}}]{Ligorini2021b}%
	  \BibitemOpen
	  \bibfield  {author} {\bibinfo {author} {\bibfnamefont {A.}~\bibnamefont {Ligorini}}, \bibinfo {author} {\bibfnamefont {J.}~\bibnamefont {Niemiec}}, \bibinfo {author} {\bibfnamefont {O.}~\bibnamefont {Kobzar}}, \bibinfo {author} {\bibfnamefont {M.}~\bibnamefont {Iwamoto}}, \bibinfo {author} {\bibfnamefont {A.}~\bibnamefont {Bohdan}}, \bibinfo {author} {\bibfnamefont {M.}~\bibnamefont {Pohl}}, \bibinfo {author} {\bibfnamefont {Y.}~\bibnamefont {Matsumoto}}, \bibinfo {author} {\bibfnamefont {T.}~\bibnamefont {Amano}}, \bibinfo {author} {\bibfnamefont {S.}~\bibnamefont {Matsukiyo}},\ and\ \bibinfo {author} {\bibfnamefont {M.}~\bibnamefont {Hoshino}},\ }\bibfield  {title} {\bibinfo {title} {{Mildly relativistic magnetized shocks in electron--ion plasmas -- II. Particle acceleration and heating}},\ }\href {https://doi.org/10.1093/mnras/stab220} {\bibfield  {journal} {\bibinfo  {journal} {Mon. Not. R. Astron. Soc.}\ }\textbf {\bibinfo {volume} {502}},\ \bibinfo {pages} {5065} (\bibinfo {year} {2021}{\natexlab{b}})}\BibitemShut {NoStop}%
	\bibitem [{\citenamefont {Matsumoto}\ \emph {et~al.}(2015)\citenamefont {Matsumoto}, \citenamefont {Amano}, \citenamefont {Kato},\ and\ \citenamefont {Hoshino}}]{Matsumoto2015}%
	  \BibitemOpen
	  \bibfield  {author} {\bibinfo {author} {\bibfnamefont {Y.}~\bibnamefont {Matsumoto}}, \bibinfo {author} {\bibfnamefont {T.}~\bibnamefont {Amano}}, \bibinfo {author} {\bibfnamefont {T.~N.}\ \bibnamefont {Kato}},\ and\ \bibinfo {author} {\bibfnamefont {M.}~\bibnamefont {Hoshino}},\ }\bibfield  {title} {\bibinfo {title} {{Stochastic electron acceleration during spontaneous turbulent reconnection in a strong shock wave}},\ }\href {https://doi.org/10.1126/science.1260168} {\bibfield  {journal} {\bibinfo  {journal} {Science}\ }\textbf {\bibinfo {volume} {347}},\ \bibinfo {pages} {974} (\bibinfo {year} {2015})}\BibitemShut {NoStop}%
	\bibitem [{\citenamefont {Matsumoto}\ \emph {et~al.}(2017)\citenamefont {Matsumoto}, \citenamefont {Amano}, \citenamefont {Kato},\ and\ \citenamefont {Hoshino}}]{Matsumoto2017}%
	  \BibitemOpen
	  \bibfield  {author} {\bibinfo {author} {\bibfnamefont {Y.}~\bibnamefont {Matsumoto}}, \bibinfo {author} {\bibfnamefont {T.}~\bibnamefont {Amano}}, \bibinfo {author} {\bibfnamefont {T.~N.}\ \bibnamefont {Kato}},\ and\ \bibinfo {author} {\bibfnamefont {M.}~\bibnamefont {Hoshino}},\ }\bibfield  {title} {\bibinfo {title} {{Electron Surfing and Drift Accelerations in a Weibel-Dominated High-Mach-Number Shock}},\ }\href {https://doi.org/10.1103/PhysRevLett.119.105101} {\bibfield  {journal} {\bibinfo  {journal} {Phys. Rev. Lett.}\ }\textbf {\bibinfo {volume} {119}},\ \bibinfo {pages} {105101} (\bibinfo {year} {2017})}\BibitemShut {NoStop}%
	\bibitem [{\citenamefont {Ikeya}\ and\ \citenamefont {Matsumoto}(2015)}]{Ikeya2015}%
	  \BibitemOpen
	  \bibfield  {author} {\bibinfo {author} {\bibfnamefont {N.}~\bibnamefont {Ikeya}}\ and\ \bibinfo {author} {\bibfnamefont {Y.}~\bibnamefont {Matsumoto}},\ }\bibfield  {title} {\bibinfo {title} {{Stability property of numerical Cherenkov radiation and its application to relativistic shock simulations}},\ }\href {https://doi.org/10.1093/pasj/psv052} {\bibfield  {journal} {\bibinfo  {journal} {Publ. Astron. Soc. Jpn.}\ }\textbf {\bibinfo {volume} {67}},\ \bibinfo {pages} {64} (\bibinfo {year} {2015})}\BibitemShut {NoStop}%
	\bibitem [{\citenamefont {Iwamoto}\ \emph {et~al.}(2019)\citenamefont {Iwamoto}, \citenamefont {Amano}, \citenamefont {Hoshino}, \citenamefont {Matsumoto}, \citenamefont {Niemiec}, \citenamefont {Ligorini}, \citenamefont {Kobzar},\ and\ \citenamefont {Pohl}}]{Iwamoto2019}%
	  \BibitemOpen
	  \bibfield  {author} {\bibinfo {author} {\bibfnamefont {M.}~\bibnamefont {Iwamoto}}, \bibinfo {author} {\bibfnamefont {T.}~\bibnamefont {Amano}}, \bibinfo {author} {\bibfnamefont {M.}~\bibnamefont {Hoshino}}, \bibinfo {author} {\bibfnamefont {Y.}~\bibnamefont {Matsumoto}}, \bibinfo {author} {\bibfnamefont {J.}~\bibnamefont {Niemiec}}, \bibinfo {author} {\bibfnamefont {A.}~\bibnamefont {Ligorini}}, \bibinfo {author} {\bibfnamefont {O.}~\bibnamefont {Kobzar}},\ and\ \bibinfo {author} {\bibfnamefont {M.}~\bibnamefont {Pohl}},\ }\bibfield  {title} {\bibinfo {title} {{Precursor Wave Amplification by Ion--Electron Coupling through Wakefield in Relativistic Shocks}},\ }\href {https://doi.org/10.3847/2041-8213/ab4265} {\bibfield  {journal} {\bibinfo  {journal} {Astrophys. J. Lett.}\ }\textbf {\bibinfo {volume} {883}},\ \bibinfo {pages} {L35} (\bibinfo {year} {2019})}\BibitemShut {NoStop}%
	\bibitem [{\citenamefont {Sobacchi}\ \emph {et~al.}(2020)\citenamefont {Sobacchi}, \citenamefont {Lyubarsky}, \citenamefont {Beloborodov},\ and\ \citenamefont {Sironi}}]{Sobacchi2020}%
	  \BibitemOpen
	  \bibfield  {author} {\bibinfo {author} {\bibfnamefont {E.}~\bibnamefont {Sobacchi}}, \bibinfo {author} {\bibfnamefont {Y.}~\bibnamefont {Lyubarsky}}, \bibinfo {author} {\bibfnamefont {A.~M.}\ \bibnamefont {Beloborodov}},\ and\ \bibinfo {author} {\bibfnamefont {L.}~\bibnamefont {Sironi}},\ }\bibfield  {title} {\bibinfo {title} {{Self-modulation of fast radio bursts}},\ }\href {https://doi.org/10.1093/mnras/staa3248} {\bibfield  {journal} {\bibinfo  {journal} {Mon. Not. R. Astron. Soc.}\ }\textbf {\bibinfo {volume} {500}},\ \bibinfo {pages} {272} (\bibinfo {year} {2020})}\BibitemShut {NoStop}%
	\bibitem [{\citenamefont {Sobacchi}\ \emph {et~al.}(2022)\citenamefont {Sobacchi}, \citenamefont {Lyubarsky}, \citenamefont {Beloborodov},\ and\ \citenamefont {Sironi}}]{Sobacchi2022}%
	  \BibitemOpen
	  \bibfield  {author} {\bibinfo {author} {\bibfnamefont {E.}~\bibnamefont {Sobacchi}}, \bibinfo {author} {\bibfnamefont {Y.}~\bibnamefont {Lyubarsky}}, \bibinfo {author} {\bibfnamefont {A.~M.}\ \bibnamefont {Beloborodov}},\ and\ \bibinfo {author} {\bibfnamefont {L.}~\bibnamefont {Sironi}},\ }\bibfield  {title} {\bibinfo {title} {{Filamentation of fast radio bursts in magnetar winds}},\ }\href {https://doi.org/10.1093/mnras/stac251} {\bibfield  {journal} {\bibinfo  {journal} {Mon. Not. R. Astron. Soc.}\ }\textbf {\bibinfo {volume} {511}},\ \bibinfo {pages} {4766} (\bibinfo {year} {2022})}\BibitemShut {NoStop}%
	\bibitem [{\citenamefont {Ghosh}\ \emph {et~al.}(2022)\citenamefont {Ghosh}, \citenamefont {Kagan}, \citenamefont {Keshet},\ and\ \citenamefont {Lyubarsky}}]{Ghosh2022}%
	  \BibitemOpen
	  \bibfield  {author} {\bibinfo {author} {\bibfnamefont {A.}~\bibnamefont {Ghosh}}, \bibinfo {author} {\bibfnamefont {D.}~\bibnamefont {Kagan}}, \bibinfo {author} {\bibfnamefont {U.}~\bibnamefont {Keshet}},\ and\ \bibinfo {author} {\bibfnamefont {Y.}~\bibnamefont {Lyubarsky}},\ }\bibfield  {title} {\bibinfo {title} {{Nonlinear Electromagnetic-wave Interactions in Pair Plasma. I. Nonrelativistic Regime}},\ }\href {https://doi.org/10.3847/1538-4357/ac581d} {\bibfield  {journal} {\bibinfo  {journal} {Astrophys. J.}\ }\textbf {\bibinfo {volume} {930}},\ \bibinfo {pages} {106} (\bibinfo {year} {2022})}\BibitemShut {NoStop}%
	\bibitem [{\citenamefont {Sobacchi}\ \emph {et~al.}(2023)\citenamefont {Sobacchi}, \citenamefont {Lyubarsky}, \citenamefont {Beloborodov}, \citenamefont {Sironi},\ and\ \citenamefont {Iwamoto}}]{Sobacchi2023}%
	  \BibitemOpen
	  \bibfield  {author} {\bibinfo {author} {\bibfnamefont {E.}~\bibnamefont {Sobacchi}}, \bibinfo {author} {\bibfnamefont {Y.}~\bibnamefont {Lyubarsky}}, \bibinfo {author} {\bibfnamefont {A.~M.}\ \bibnamefont {Beloborodov}}, \bibinfo {author} {\bibfnamefont {L.}~\bibnamefont {Sironi}},\ and\ \bibinfo {author} {\bibfnamefont {M.}~\bibnamefont {Iwamoto}},\ }\bibfield  {title} {\bibinfo {title} {{Saturation of the Filamentation Instability and Dispersion Measure of Fast Radio Bursts}},\ }\href {https://doi.org/10.3847/2041-8213/acb260} {\bibfield  {journal} {\bibinfo  {journal} {Astrophys. J. Lett.}\ }\textbf {\bibinfo {volume} {943}},\ \bibinfo {pages} {L21} (\bibinfo {year} {2023})}\BibitemShut {NoStop}%
	\bibitem [{\citenamefont {Iwamoto}\ \emph {et~al.}(2023)\citenamefont {Iwamoto}, \citenamefont {Sobacchi},\ and\ \citenamefont {Sironi}}]{Iwamoto2023}%
	  \BibitemOpen
	  \bibfield  {author} {\bibinfo {author} {\bibfnamefont {M.}~\bibnamefont {Iwamoto}}, \bibinfo {author} {\bibfnamefont {E.}~\bibnamefont {Sobacchi}},\ and\ \bibinfo {author} {\bibfnamefont {L.}~\bibnamefont {Sironi}},\ }\bibfield  {title} {\bibinfo {title} {{Kinetic simulations of the filamentation instability in pair plasmas}},\ }\href {https://doi.org/10.1093/mnras/stad1100} {\bibfield  {journal} {\bibinfo  {journal} {Mon. Not. R. Astron. Soc.}\ }\textbf {\bibinfo {volume} {522}},\ \bibinfo {pages} {2133} (\bibinfo {year} {2023})}\BibitemShut {NoStop}%
	\bibitem [{\citenamefont {Katz}(2022{\natexlab{b}})}]{Katz2022a}%
	  \BibitemOpen
	  \bibfield  {author} {\bibinfo {author} {\bibfnamefont {J.~I.}\ \bibnamefont {Katz}},\ }\bibfield  {title} {\bibinfo {title} {{The environment and constraints on the mass of FRB 190520B}},\ }\href {https://doi.org/10.1093/mnrasl/slac051} {\bibfield  {journal} {\bibinfo  {journal} {Mon. Not. R. Astron. Soc.}\ }\textbf {\bibinfo {volume} {514}},\ \bibinfo {pages} {L27} (\bibinfo {year} {2022}{\natexlab{b}})}\BibitemShut {NoStop}%
	\bibitem [{\citenamefont {Lyubarsky}(2006)}]{Lyubarsky2006}%
	  \BibitemOpen
	  \bibfield  {author} {\bibinfo {author} {\bibfnamefont {Y.}~\bibnamefont {Lyubarsky}},\ }\bibfield  {title} {\bibinfo {title} {{Electron‐Ion Coupling Upstream of Relativistic Collisionless Shocks}},\ }\href {https://doi.org/10.1086/508606} {\bibfield  {journal} {\bibinfo  {journal} {Astrophys. J.}\ }\textbf {\bibinfo {volume} {652}},\ \bibinfo {pages} {1297} (\bibinfo {year} {2006})}\BibitemShut {NoStop}%
	\bibitem [{\citenamefont {Hoshino}(2008)}]{Hoshino2008}%
	  \BibitemOpen
	  \bibfield  {author} {\bibinfo {author} {\bibfnamefont {M.}~\bibnamefont {Hoshino}},\ }\bibfield  {title} {\bibinfo {title} {{Wakefield Acceleration by Radiation Pressure in Relativistic Shock Waves}},\ }\href {https://doi.org/10.1086/523665} {\bibfield  {journal} {\bibinfo  {journal} {Astrophys. J.}\ }\textbf {\bibinfo {volume} {672}},\ \bibinfo {pages} {940} (\bibinfo {year} {2008})}\BibitemShut {NoStop}%
	\bibitem [{\citenamefont {Rybicki}\ and\ \citenamefont {Lightman}(1979)}]{Rybicki1979}%
	  \BibitemOpen
	  \bibfield  {author} {\bibinfo {author} {\bibfnamefont {G.~B.}\ \bibnamefont {Rybicki}}\ and\ \bibinfo {author} {\bibfnamefont {A.~D.}\ \bibnamefont {Lightman}},\ }\href@noop {} {\emph {\bibinfo {title} {Radiative Processes in Astrophysics}}}\ (\bibinfo  {publisher} {John Wiley \& Sons, Inc.},\ \bibinfo {address} {New York},\ \bibinfo {year} {1979})\BibitemShut {NoStop}%
	\bibitem [{\citenamefont {Babul}\ and\ \citenamefont {Sironi}(2020)}]{Babul2020}%
	  \BibitemOpen
	  \bibfield  {author} {\bibinfo {author} {\bibfnamefont {A.-N.}\ \bibnamefont {Babul}}\ and\ \bibinfo {author} {\bibfnamefont {L.}~\bibnamefont {Sironi}},\ }\bibfield  {title} {\bibinfo {title} {{The synchrotron maser emission from relativistic magnetized shocks: dependence on the pre-shock temperature}},\ }\href {https://doi.org/10.1093/mnras/staa2612} {\bibfield  {journal} {\bibinfo  {journal} {Mon. Not. R. Astron. Soc.}\ }\textbf {\bibinfo {volume} {499}},\ \bibinfo {pages} {2884} (\bibinfo {year} {2020})}\BibitemShut {NoStop}%
	\bibitem [{\citenamefont {Iwamoto}\ \emph {et~al.}(2022)\citenamefont {Iwamoto}, \citenamefont {Amano}, \citenamefont {Matsumoto}, \citenamefont {Matsukiyo},\ and\ \citenamefont {Hoshino}}]{Iwamoto2022}%
	  \BibitemOpen
	  \bibfield  {author} {\bibinfo {author} {\bibfnamefont {M.}~\bibnamefont {Iwamoto}}, \bibinfo {author} {\bibfnamefont {T.}~\bibnamefont {Amano}}, \bibinfo {author} {\bibfnamefont {Y.}~\bibnamefont {Matsumoto}}, \bibinfo {author} {\bibfnamefont {S.}~\bibnamefont {Matsukiyo}},\ and\ \bibinfo {author} {\bibfnamefont {M.}~\bibnamefont {Hoshino}},\ }\bibfield  {title} {\bibinfo {title} {{Particle Acceleration by Pickup Process Upstream of Relativistic Shocks}},\ }\href {https://doi.org/10.3847/1538-4357/ac38aa} {\bibfield  {journal} {\bibinfo  {journal} {Astrophys. J.}\ }\textbf {\bibinfo {volume} {924}},\ \bibinfo {pages} {108} (\bibinfo {year} {2022})}\BibitemShut {NoStop}%
	\end{thebibliography}
%

\end{document}